\documentclass[aps,prl,twocolumn,floatfix,longbibliography,nofootinbib,tbtags]{revtex4-1}
\usepackage[utf8]{inputenc}
\usepackage{natbib}
\usepackage{graphicx}
\usepackage{xcolor}
\usepackage{mathtools}
\usepackage{amsmath}
\usepackage[colorlinks, linkcolor=blue]{hyperref}
\hypersetup{colorlinks,allcolors=black}
\usepackage{amssymb}
\usepackage{gensymb}
\usepackage{float}
\usepackage{txfonts}
\usepackage{tabularx,graphicx}
\usepackage{epstopdf}
\usepackage{latexsym}
\usepackage{color, colortbl}
\usepackage{psfrag}
\usepackage{bbm}
\usepackage{bm}
\usepackage{titlesec}
\usepackage{dsfont}
\usepackage{feynmp}
\usepackage{slashed}
\usepackage{multirow}
\usepackage[normalem]{ulem}
\renewcommand{\vec}[1]{\boldsymbol{#1}}

\def \r {{\vec r}}

\def \S {{\cal{S}}}

\def \beq {\begin{eqnarray}}
\def \eeq {\end{eqnarray}}
\def \tn {\textnormal}

\def \N {{\cal{N}}}

\begin{document}
\title{Continuous {Wigner-}Mott Transitions at $\nu=1/5$}
\author{Thomas G. Kiely}
\author{Debanjan Chowdhury}
\affiliation{Department of Physics, Cornell University, Ithaca, New York 14853, USA.}

\begin{abstract}
Electrons can organize themselves into charge-ordered states to minimize the effects of long-ranged Coulomb interactions. In the presence of a lattice, commensurability constraints lead to the emergence of incompressible {Wigner-}Mott insulators at various rational electron fillings, $\nu~=p/q$. The mechanism for quantum fluctuation-mediated melting of the Mott insulators with increasing electron kinetic energy remains an outstanding problem. Here, using matrix product state techniques, we analyze 
the bandwidth-tuned transition out of the {Wigner-}Mott insulator at $\nu=1/5$ in an extended Hubbard model on infinite cylinders of varying circumference. For the two-leg ladder, the transition from the Mott insulator to the Luttinger liquid proceeds via a distinct intermediate phase with gapless Cooper-pairs and gapped electronic excitations. The resulting Luther-Emery liquid is the analog of a strongly fluctuating superconductor. We place these results in the context of a low-energy bosonization based theory for the transition. On the five-leg cylinder, we provide numerical evidence for a direct continuous transition between the {Wigner-}Mott insulator and a metallic phase across which the spin and charge-gaps vanish simultaneously. We comment on the connections to ongoing experiments in dual-gated bilayer moir\'e transition metal dichalcogenide materials. 
\end{abstract}

\maketitle

{\it Introduction.-} The emergence of {Wigner-}Mott insulators at a partial commensurate filling of electronic bands is one of the hallmarks of an interaction-induced phenomenon \cite{mott1990}. Despite being an old problem, much remains to be understood about how a {Wigner-}Mott {(WM)} insulator emerges from a Landau-Fermi liquid (FL) at a fixed electronic filling ($\nu$) as a function of increasing strength of interactions. The FL metal hosts an electronic Fermi surface whose area is fixed by Luttinger's theorem \cite{LW,oshikawa}. On the other hand, the {WM} insulator does {\it not} host any electronic Fermi surface and spontaneously breaks space-group (and, possibly, spin-rotation) symmetries. Given these differences, one might expect that the most common scenario would be for the metal-insulator transition to be first-order in nature, which is seen in many solid-state materials \cite{imada1998}.
An alternative ``weak-coupling'' perspective suggests that the transition can proceed via intermediate metallic phases with broken translational symmetry and an even number of electrons in the enlarged unit-cell. Examples of both classes of transitions, including the additional effects of disorder, have been analyzed in a large body of earlier work \cite{jamei_universal_2005,camjayi_coulomb_2008,amaricci_extended_2010,vu}. The most intriguing scenario involves a direct continuous transition between a symmetry-preserving FL metal and a {WM} insulator. As a matter of principle, such continuous transitions can be described using quantum field theoretic methods involving fractionalized degrees of freedom and emergent gauge-fields \cite{musser2022continuous,Xu22}, but they typically rely on artificial limits to make computational progress. 

\begin{figure}
    \centering
    \includegraphics[width=3.375in]{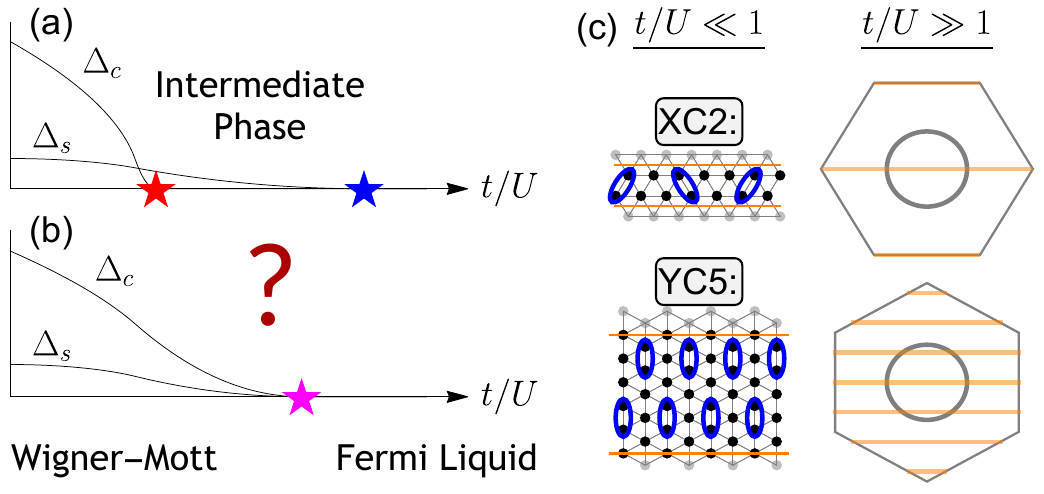}
    \caption{Schematic for two possible scenarios for continuous bandwidth-tuned Mott transitions at $\nu=1/5$, where $\Delta_c,~\Delta_s$ vanish at: (a) two separate QCPs, with an intermediate gapless phase, or (b) the same QCP but with distinct critical exponents. (c) The two quasi-2D infinite cylinder geometries, XC2 and YC5, along with the respective Mott insulators ($t/U\ll 1$) and the electronic Fermi surface in the metallic phase ($t/U\gg 1$). The blue ovals denote singlet bonds and the orange lines denote cuts through the electronic Fermi surface, corresponding to the allowed momentum modes around the cylinder circumference.}
    \label{fig:introFig}
\end{figure}

In this letter, we study the transition(s) between a {WM} insulator and a FL metal at a fixed filling $\nu=1/5$ for spinful electrons on the triangular lattice using infinite matrix product state (MPS) techniques \cite{RMP_MPS,SCHOLLWOCKreview}. 
In the strong-coupling regime, superexchange leads to a spin-singlet Mott-insulating ground state with finite spin and charge gaps, $\Delta_s,~\Delta_c$, respectively; see Fig.~\ref{fig:introFig}. 
Starting with the fully-gapped Mott insulator, which can be efficiently represented using MPS, we address the quantum fluctuation-induced melting 
with increasing single-electron bandwidth. In particular, are there intermediate gapless phases distinct from a symmetry-preserving
FL metal (Fig.~\ref{fig:introFig}a), or is there a direct transition to the FL across which $\Delta_s,~\Delta_c$ vanish simultaneously (Fig.~\ref{fig:introFig}b)? 

With increasing quantum fluctuations, the spin-singlets in the {Wigner-}Mott insulator can melt into delocalized, 
strongly fluctuating Cooper-pairs before breaking apart to reveal the electronic excitations. For a particular set of microscopic interaction parameters, we find compelling evidence for this two-step transition on two-leg ladders where the intervening Luther-Emery liquid \cite{luther1974,SKreview} hosts a gap to single-electron excitations, but not to the spin-singlet Cooper pairs. We emphasize that this is a ``strong-coupling" route to engineering superconductivity in a model with purely repulsive interactions, without the need for any microscopic attraction.
This curious result derives from the {Wigner-}Mott state, which can be understood as a crystal of localized Cooper-pairs (see Fig.~\ref{fig:introFig}c). 
The recent report of superconductivity in twisted WSe$_2$ obtained by melting a correlated insulator \cite{TMDSC} at $\nu=1$ is an interesting experimental example of similar phenomenology; see also Ref.~\cite{Dean24}. Notably, the conditions under which these pairing correlations emerge (at low-density and with long-range interactions) are markedly distinct from prior reports of pairing correlations at weak coupling~\cite{fabrizio1993,balents1996} or at unit filling~\cite{dagotto1996,kuroki,zhou2023} in two-leg ladders, and they differ from recent theoretical predictions under comparable conditions~\cite{musser2022}.

On the five-leg cylinder and for a similar choice of parameters, we find evidence for a {\it direct continuous transition} between the {Wigner-}Mott insulator with broken translational symmetry and a gapless metallic phase. 
Such a transition is seemingly at odds with any Landau-Ginzburg-Wilson-based paradigm for continuous quantum phase transitions, as there is no {\it a priori} reason for the spin and charge orders to vanish and the electronic Fermi surface to appear simultaneously. This result is corroborated in part by a non-trivial scaling collapse associated with the extracted spin and charge ``gaps", respectively.

{\it Experimental motivation.-} Bilayers of transition metal dichalcogenides (TMDs) realize an effective moir{\'e} triangular lattice, and have been used to study a bandwidth-tuned continuous metal-Mott insulator transition at $\nu=1/2$~\cite{li2021continuous,ghiotto2021quantum}. In parallel, a number of experiments using moir{\'e} TMDs have reported evidence for a plethora of {Wigner-}Mott insulators at electron fillings $\nu=1/6,~1/5,~1/4$ etc., induced by the screened long-range Coulomb interactions ~\cite{xu2020correlated,regan_mott_2020,huang_correlated_2021}. Wigner crystals in the absence of a moir\'e potential have been reported across different platforms in earlier work \cite{QHWC1,QHWC2,QHWC3,TMDWC1,TMDWC2}. 
While the nature of the electron spin-configuration, and its possible ordering, is presently 
unclear for all of these {WM} insulators, the charge-order for $\nu=1/6$ has been imaged directly \cite{li_imaging_2021}. The competing spin-exchange interactions
are highly frustrated and the ordering (or lack thereof) is a delicate question
\cite{roger,SK99,spivak}. In the context of moir\'e systems, previous theoretical effort has focused on the crystalline regime deep in the Mott insulator \cite{PP21,LF21}; Hartree-Fock  \cite{SDS20,pan_interaction-driven_2021,SDS22}, classical Monte-Carlo \cite{matty}, and DMRG-based \cite{Zhou22} methods have also been used to study the competition between spin and charge-orderings over a range of coupling strength and density. Momentum-space-based exact diagonalization methods have also been employed to study the metal-insulator transition for a host of other fillings \cite{morales-duran_metal-insulator_2021,morales22}. The Mott insulator \footnote{The same filling is denoted $\nu_c=2/5$ in Ref.~\cite{xu2020correlated}, measured relative to the full filling of the band ($\nu_c=2$).} at $\nu=1/5$ provides a useful starting point to study the onset of electron delocalization and melting of the spin-gap, going beyond any Hartree-Fock or classical Monte-Carlo-based approach.

{\it Model \& Method.-} We study the ground state phase diagram of the extended Hubbard model on the triangular lattice, given by
\begin{subequations}
\beq
    H&=&H_{\tn{kin}} + H_{\tn{int}},\\
    H_{\tn{kin}} &=&-t\sum_{\langle \r,\r'\rangle,\sigma}\left( c^\dagger_{\r\sigma}c^{\phantom\dagger}_{\r'\sigma}+{\rm H.c.}\right) - \mu\sum_\r n_\r,\\
    H_{\tn{int}} &=& U\sum_\r n_{\r\uparrow}n_{\r\downarrow} +\frac{1}{2}\sum_{\r\neq \r'}V(\r-\r')~n_{\r}n_{\r'}.
\eeq
\end{subequations}
Here the electron creation and annihilation operators at site $\r$ with spin $\sigma$ are denoted $c^\dagger_{\r\sigma},~c_{\r\sigma}$, respectively. The on-site interaction, $U$, and further neighbor interactions, $V(\r)$, are kept fixed, with the latter determined by the screened Coulomb interaction in bilayer TMD experiments~\cite{xu2020correlated,li2021continuous,ghiotto2021quantum}. For our calculations, we truncate $V(\r)$ at fourth-nearest-neighbor interactions on the triangular lattice with $V_2/V_1\approx 0.512$, $V_3/V_1\approx 0.423$, and $V_4/V_1\approx 0.284$, where $V_n$ is the $n^{\tn{th}}$ nearest-neighbor interaction strength \cite{si}. We choose $V_1/U=0.5$. The single electron hopping is $t$, and the chemical potential, $\mu$, couples to the total electron density with $n_\r=\sum_\sigma c^\dagger_{\r\sigma}c^{\phantom\dagger}_{\r\sigma}$. In the remainder of this manuscript, we focus on the electron filling fraction $\nu=1/5$ and in the zero-magnetization sector. The phase diagram is then studied by varying $t$ at fixed filling and interaction. 

We make use of infinite matrix product state (iMPS) techniques on two quasi-2D geometries: the two-leg ladder (XC2) and the five-leg  cylinder (YC5). 
The geometries are shown schematically in Fig.~\ref{fig:introFig}(c). We numerically determine variational ground states as function of the iMPS bond dimension, $\chi$~\cite{vumps,vumps2}. Wherever possible, we extrapolate our results in the limit $\chi\to\infty$. See~\cite{si} for further details.

\begin{figure*}
    \centering
    \includegraphics[width=17.5cm]{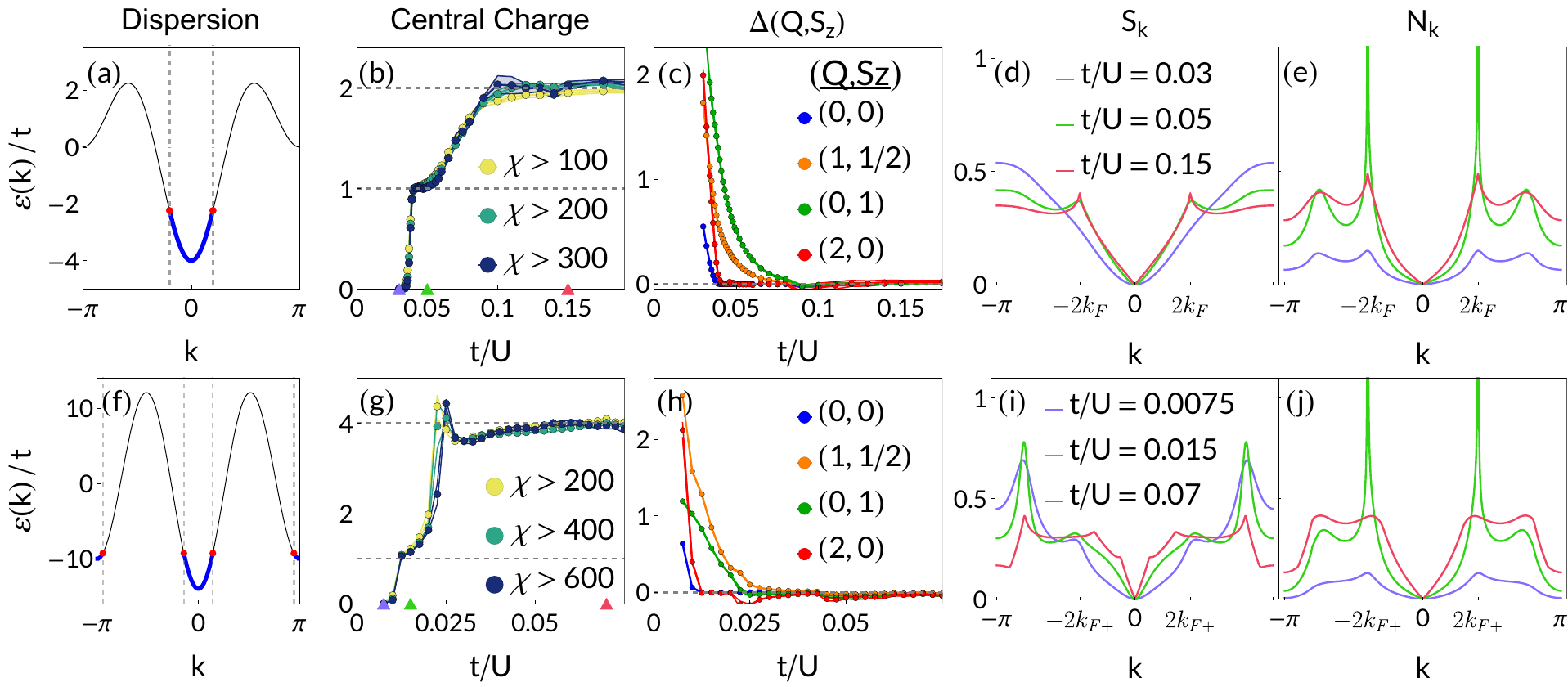}
    \caption{The two-step Mott-metal transition, as shown in Fig.~\ref{fig:introFig}a, on the XC2 geometry. Panels (b)-(e) support the scenario $C0S0\rightarrow C1S0\rightarrow C1S1$ for the dispersion shown in (a) with $2$ Fermi points. Similarly, panels (g)-(h) support the scenario $C0S0\rightarrow C1S0\rightarrow C2S2$ for the dispersion shown in (f) with $4$ Fermi points. The $C1S0$ phase is a Luther-Emery liquid. Conformal central charge ($C$) shows a sharp transition from $C=0$ to $C=1$ across (b) $t_{c1}/U=0.04$ (for (a)), and (g) $t_{c1}/U=0.01$ (for (f)). There is a subsequent gradual increase from (b) $C=1$ to $C=2$ across $t_{c2}/U\approx 0.1$ (for (a)), and (g) $C=1$ to $C=4$ across $t_{c2}/U\approx0.024$ (for (f)). Different colors represent $C$ obtained by removing varying number of low-$\chi$ states (see main text). (c) and (h) Extrapolated spectral gaps, $\Delta(Q,S_z;\chi)$, in different symmetry sectors across the phase diagram for the dispersions in (a) and (f), respectively. Below $t_{c1}$ all symmetry sectors are gapped; between $t_{c1}$ and $t_{c2}$ the spin sectors remain gapped while the $(Q,S_z)=(0,0)$ and $(2,0)$ gaps vanish. For $t>t_{c2}$ all sectors are gapless. (d),(i) Spin and (e),(j) density structure factors for three representative points in the phase diagram (marked by the `$\blacktriangle$' in (b) and (g)). In a gapless sector associated with a specific $\{Q,~S_z\}$, the corresponding SF$\sim |k|$ for small $k$ and develops singular cusps at $2k_F$, or $2k_{F+}=2(k_{F1}+k_{F2})$ (see main text).}
    \label{fig:twolegpanel}
\end{figure*}
{\it Two-leg ladders.-}  We perform calculations with the XC2 geometry for two different choices of the hopping parameters.
We denote the hopping amplitude along the long direction $t$ and the amplitude along the short direction $t^\prime$ (see Fig.~\ref{fig:introFig}). For 
(i) $t=t^\prime$, the bandstructure leads to $2$ Fermi points (FP), as in Fig.~\ref{fig:twolegpanel}a, while for (ii) $t^\prime=6t$ there are $4$ FP as in Fig.~\ref{fig:twolegpanel}f.
Let us denote the metallic phase in the corresponding bosonized model as $C\alpha S\beta$, where $\alpha~(\beta)$ denote the number of charge (spin) modes, respectively. For both cases, we find a Mott insulator on the two-leg ladder for 
$U/t\gg1$ and a metallic Luttinger liquid (LL) for
$U/t\ll1$ (see Fig.~\ref{fig:introFig}c). 
We do {\it not}, however, find a direct transition between these phases -- rather, with increasing $t/U$, the {WM} insulator first transitions into an intermediate phase with {\it gapless} spin-singlet Cooper-pair excitations and a gap to spin excitations.
Further increasing $t$, there is a subsequent transition at $t_{c2}/U$ where the spin gap closes and we recover the LL. Thus, the melting of the {WM} insulator fits into the schematic shown in Fig.~\ref{fig:introFig}a. 
For case (i), the Mott insulator melting transition follows the sequence $C0S0\rightarrow C1S0\rightarrow C1S1$, while for case (ii), the sequence is given by $C0S0\rightarrow C1S0\rightarrow C2S2$. In both cases, the transition proceeds via an intermediate Luther-Emery liquid ($C1S0$). This is the fluctuating ``superconducting" phase, that appears in a purely repulsive model without {\it any} retardation effects by melting the localized spin-singlet Cooper pairs. 
This two-step transition is summarized in Fig.~\ref{fig:twolegpanel} for cases (i) (panels a-e) and (ii) (panels f-j). 
As the phenomenology is largely the same, we will limit our discussion in the main text to case (i); please see~\cite{si} for a discussion of case (ii) as well as further details on these observables.

We use three diagnostics to study the metal-insulator transition: the conformal central charge, the inverse correlation length, and the static structure factors. The central charge, $C$, is shown as a function of $t/U$ in Fig.~\ref{fig:twolegpanel}b. In the Mott phase, the ground state is gapped and $C=0$. Across $t_{c1}/U\approx 0.04$ we see a sharp increase in $C$ followed by a plateau around $C\approx 1$. This plateau becomes more prominent at larger bond dimensions (different colors in Fig.~\ref{fig:twolegpanel}b). The central charge $C=1$ is consistent with two gapless modes, while the two-component LL has four gapless modes. Hence, this plateau is strongly suggestive of a distinct intermediate phase. Across $t_{c2}/U\approx0.1$, the system enters the LL phase with $C=2$. 
In~\cite{si} we provide an estimate of $t_{c1}$ and $t_{c2}$, with standard error bars, using a scaling collapse.

The inverse correlation lengths are shown in Fig.~\ref{fig:twolegpanel}c.
The iMPS variational wavefunction is characterized by a spectrum of correlation lengths, $\{\xi(\chi)\}$, which has been shown to map onto the low-energy excitation spectrum (up to an overall scale factor) as $\chi\to\infty$~\cite{Zauner2015,eberharter2023}. In Fig.~\ref{fig:twolegpanel}c, we show the dominant inverse correlation lengths (``gaps") in some of the relevant symmetry sectors, labeled by ${\rm U}(1)$ charge ($Q$) and spin ($S_z$) quantum numbers~\cite{szasz2020}: $\Delta(Q,S_z)\equiv \lim_{\chi\to\infty} 1/\xi(Q,S_z;\chi)$~\cite{si}. In the Mott phase, all symmetry sectors are gapped. As we cross $t_{c1}$, the gaps for spinless excitations ($(0,0)$ and $(2,0)$) vanish while those for spinful excitations ($(1,1/2)$ and $(0,1)$) are finite. This is consistent with $C=1$ and reveals that the intermediate phase has gapless charge excitations with a gap to spin excitations. Across $t_{c2}$, all symmetry sectors show a vanishing gap, consistent with the expected behavior in the LL phase.

The density and spin structure factors (SF) are shown in Figs.~\ref{fig:twolegpanel}d and e.
The spin SF is defined as $\S_k =\frac{1}{5}\sum_{j=1}^5\sum_l e^{ilk}\langle {\vec S}_j\cdot {\vec S}_{l+j}\rangle$ where $j$ iterates over the 5-site unit cell and 
${\vec S}_j=(1/2)\sum_{\alpha,\beta}c^\dagger_{j,\alpha}{\vec \sigma}_{\alpha,\beta}c_{j,\beta}$. We define the density SF as $\N_k = \frac{1}{5}\sum_{j=1}^5\sum_l e^{ilk}\langle (n_j-\langle n_j\rangle)(n_{l+j}-\langle n_{l+j}\rangle) \rangle$, subtracting off the average density on each site~\cite{si}. In the Mott insulator, both the density and spin SF are featureless (non-singular at any $k$). At small $k$, they vanish smoothly as $\N_k,~\S_k\sim k^2$, indicating a finite spin and charge gap, respectively~\cite{mishmash2015}. In the intermediate phase, the spin SF remains featureless while $\N_k\sim |k|$ for small $k$, indicating that the charge gap has closed. The density SF also develops singular peaks at $\pm2k_F$ that appear to diverge in the limit $\chi\to\infty$. In the LL phase, both the density and spin SF scale as $\N_k,~\S_k\sim |k|$ for small $k$ and develop singular cusps at $\pm 2k_F$~\cite{si}. 

\begin{figure}
    \centering
    \includegraphics[width=\columnwidth]{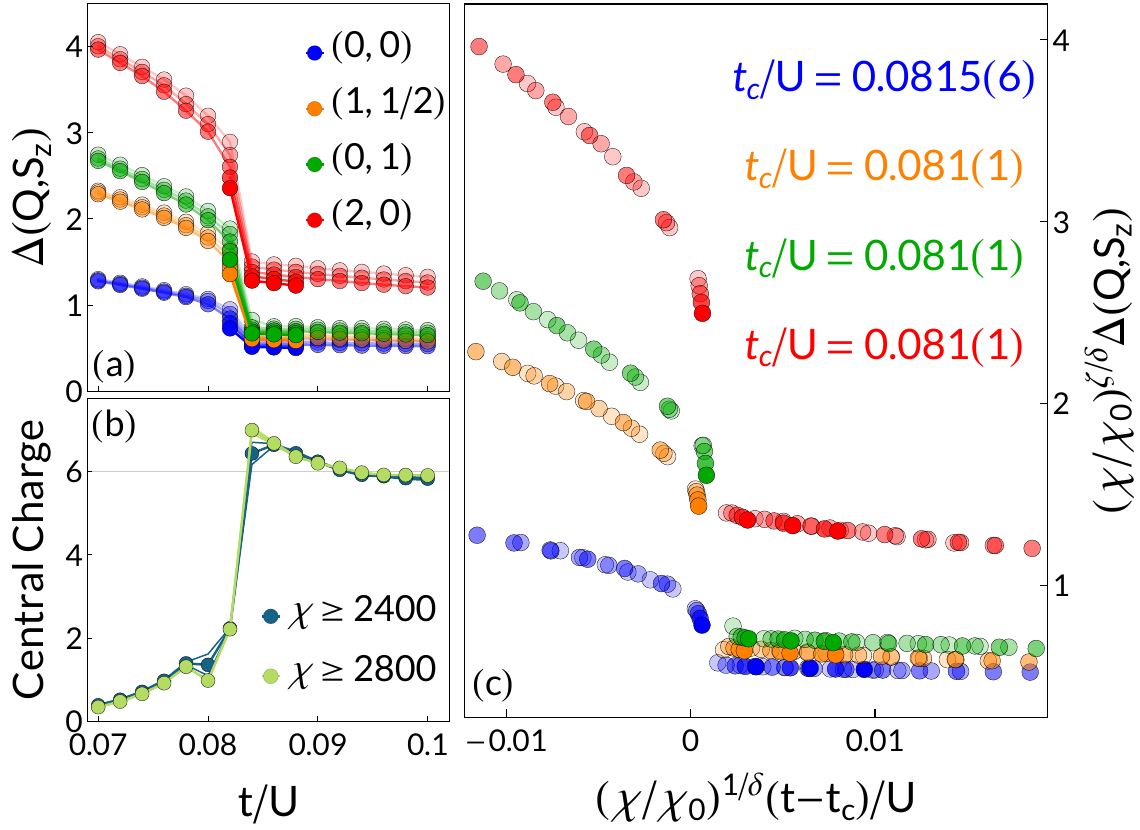}
    \caption{Mott insulator-metal transition at $\nu=1/5$ on the YC5 geometry. Panel (a) shows dominant inverse correlation lengths in different symmetry sectors, $\Delta(Q,S_z)$, as a function of $t/U$. Colors denote different sectors while the opacity denotes the bond dimension; data here is for $\chi=2400,2800,3200,3600,4000$. We see that all symmetry sectors exhibit similar behavior, showing a gap for $t/U\lesssim 0.08$ and vanishing smoothly for $t/U\gtrsim 0.084$. Panel (b) shows the central charge, extracted from a linear fit to the entanglement entropy versus correlation length, in the same region. The gapless region has a central charge $C=6$ on the YC5 geometry, consistent with panel (a), while $C$ vanishes in the gapped region. Panel (c) shows a collapse of the data in panel (a) after rescaling both axes by powers of the bond dimension (for scale, we take $\chi_0=3200$). This non-trivial result suggests that the transition is continuous, and allows us to extract critical values of $t_c/U$ within each symmetry sector (shown on plot, see main text for discussion).
    }
    \label{fig:yc5}
\end{figure}

{\it Five-leg cylinder.-} 
To study the effect of an increasing cylinder width, we perform calculations on the YC5 geometry with isotropic hopping amplitudes. Our results are summarized in Fig.~\ref{fig:yc5}. 
We show both the symmetry-resolved spectrum of inverse correlation lengths (panel (a)), derived from eigenvalues of the MPS transfer matrix, as well as the conformal central charge (panel (b)) across the transition. Data is shown for bond dimensions $\chi=2400,2800,3200$ for all data points; in the vicinity of the transition, we show additional data at $\chi=3600,4000$. We omit data below $\chi=2400$ because it shows strong finite-$\chi$ effects~\cite{si}.

We find evidence of a {Wigner-}Mott insulating phase for $t/U\lesssim 0.08$ which is well-characterized by the cartoon in Fig.~\ref{fig:introFig}c: spatially-separated pairs of electrons that form spin-singlets. This state is characterized by a spin and charge gap, as evidenced by a sharp increase in the inverse correlation lengths across all sectors.
For $t/U\gtrsim 0.084$, we find that the inverse correlation lengths in all symmetry sectors appear to vanish uniformly with increasing $\chi$ and the central charge (Fig.~\ref{fig:yc5}c) approaches $C=6$, as expected in the gapless FL phase. 
We emphasize that these inverse correlation lengths are not extrapolated with respect to bond dimension, and hence do not vanish in the FL phase (c.f. Fig.~\ref{fig:twolegpanel}). The central charge of 6 is the maximum number of gapless modes for free fermions on the YC5 cylinder, and hence provides strong evidence that we are in the FL phase.

For all $\chi$, the {Wigner-}Mott insulator melts as $t/U$ is increased; the location at which it melts, however, exhibits some $\chi$ dependence. Moreover, we are unable to apply the extrapolation method used in the XC2 geometry~\cite{si}; this is likely due to a combination of degeneracies on the cylinder geometry as well as finite-$\chi$ effects.
Intriguingly, however, we find that the gaps (inverse correlation lengths) exhibit a scaling collapse when {\it both} $\Delta(Q,S_z)$ and the value of $t/U$ are rescaled by powers of the bond dimension. This hallmark signature of a quantum critical point allows us to approximate both $t_c/U$ as well as the critical exponents. Specifically, we assume that the gaps take the scaling form $\Delta(t_c-t,\chi)\propto \chi^{\zeta/\delta} f((t_c-t)/\chi^{1/\delta})$ near the transition, which implies that $\Delta(t_c-t)\propto(t_c-t)^\zeta$ as $\chi\to\infty$. In a field-theoretic setting, the exponent $\zeta=\nu z$, can be naturally expressed in terms of the correlation length ($\nu$) and dynamical exponents ($z$), respectively, assuming that the transition is truly continuous. We then fit the parameters $\{t_c,\zeta,\delta\}$ within each independent symmetry sector, obtaining an approximate location for the transition as well as the critical exponents. The scaling collapses are shown in Fig.~\ref{fig:yc5}c. We fit each symmetry sector independently, finding very close agreement with $t_c/U=0.081(1)$. Best-fit parameters for the individual sectors are displayed on Fig.~\ref{fig:yc5}c and are reported in \cite{si}.
Fits are performed by defining a cost function in terms of the residuals from an interpolated scaling function, and error bars come from integrating the associated probability distribution~\cite{si,mortensen2005}.
Smaller error bars on $\{t_c,\zeta,\delta\}$ could be obtained by taking more data on a finer grid, but a more precise location for the transition would likely require going beyond $\chi=4000$.

{\it Discussion.-} In this letter, we have analyzed the quantum melting of a crystalline {Wigner-}Mott insulator with well formed spin-singlets into a symmetry-preserving metal using iMPS-based methods. Experiments in moir\'e TMDs are already well-placed to study such transitions in detail in the future. For two-leg ladders, we find clear evidence for an intermediate gapless Luther-Emery phase, which is distinct from the usual Luttinger liquid. {This one-dimensional analog of a fluctuating superconductor may enjoy enhanced stability due to the low dimensionality of the system -- a true 2D analog would be an interesting subject of future work. In the system studied here,} 
kinetic energy-driven quantum fluctuations 
favor melting the spin-singlets into a fluctuating (spin-gapped) superconductor without a gap to Cooper-pair excitations. Extending this picture to the case of the recent experiments \cite{TMDSC,Dean24}, which observe superconductivity in the vicinity of a correlated insulator, remains an interesting future direction. 

We have studied the effect of increasing the spatial extent along the second dimension by analyzing the same problem on five-leg cylinders.
There, we find that the intermediate gapless phase vanishes altogether, yielding to a continuous bandwidth-tuned metal-insulator transition. Developing complementary analytical methods in two spatial dimensions, going beyond the usual large-$N$ parton-based approaches, to study these transitions is clearly also desirable in light of our findings. It is especially challenging to describe such continuous metal-insulator transitions in the absence of {\it any} remnant Fermi surface in the insulating state \cite{LZDC,YZSS}, associated with even neutral (e.g. spinon) excitations \cite{senthil2008}, as observed in our five-leg cylinders. Investigating the effect of inhomogeneities on transport near this metal-insulator transition, building on previous studies at other fillings \cite{VD22,SDSdis,SKDC23} will be of direct experimental interest. Since there is a 
tendency towards superconductivity near the melting transition associated with the {Wigner-}Mott insulator, it is natural to address the fate of the ground state with additional doped holes \cite{RVB,LNW}. Given the proximity to charge-ordered states, the resulting superconductor might also be a pair-density wave \cite{PDW}. The importance of proximity to the quantum critical point(s), if any, on the associated phenomenology for the doped case also remains an important open question.

{\it Acknowledgements.-} DC thanks S. Kim, K.F. Mak, S. Musser, S. Sachdev, T. Senthil, J. Shan and L. Zou for numerous inspiring discussions and collaborations related to continuous metal-insulator transitions. TK thanks E. Mueller for help with the development and implementation of the numerical methods used in this manuscript. We thank S. Musser for clarifying discussions related to the Mott transition on the XC2 geometry, and for useful comments on an earlier version of this manuscript. 
TK is supported by an NSF grant (PHY-2110250). This work is supported in part by a CAREER grant from the NSF to DC (DMR-2237522).

\bibliography{bib}

\begin{thebibliography}{77}%
\makeatletter
\providecommand \@ifxundefined [1]{%
 \@ifx{#1\undefined}
}%
\providecommand \@ifnum [1]{%
 \ifnum #1\expandafter \@firstoftwo
 \else \expandafter \@secondoftwo
 \fi
}%
\providecommand \@ifx [1]{%
 \ifx #1\expandafter \@firstoftwo
 \else \expandafter \@secondoftwo
 \fi
}%
\providecommand \natexlab [1]{#1}%
\providecommand \enquote  [1]{``#1''}%
\providecommand \bibnamefont  [1]{#1}%
\providecommand \bibfnamefont [1]{#1}%
\providecommand \citenamefont [1]{#1}%
\providecommand \href@noop [0]{\@secondoftwo}%
\providecommand \href [0]{\begingroup \@sanitize@url \@href}%
\providecommand \@href[1]{\@@startlink{#1}\@@href}%
\providecommand \@@href[1]{\endgroup#1\@@endlink}%
\providecommand \@sanitize@url [0]{\catcode `\\12\catcode `\$12\catcode
  `\&12\catcode `\#12\catcode `\^12\catcode `\_12\catcode `\%12\relax}%
\providecommand \@@startlink[1]{}%
\providecommand \@@endlink[0]{}%
\providecommand \url  [0]{\begingroup\@sanitize@url \@url }%
\providecommand \@url [1]{\endgroup\@href {#1}{\urlprefix }}%
\providecommand \urlprefix  [0]{URL }%
\providecommand \Eprint [0]{\href }%
\providecommand \doibase [0]{http://dx.doi.org/}%
\providecommand \selectlanguage [0]{\@gobble}%
\providecommand \bibinfo  [0]{\@secondoftwo}%
\providecommand \bibfield  [0]{\@secondoftwo}%
\providecommand \translation [1]{[#1]}%
\providecommand \BibitemOpen [0]{}%
\providecommand \bibitemStop [0]{}%
\providecommand \bibitemNoStop [0]{.\EOS\space}%
\providecommand \EOS [0]{\spacefactor3000\relax}%
\providecommand \BibitemShut  [1]{\csname bibitem#1\endcsname}%
\let\auto@bib@innerbib\@empty
\bibitem [{\citenamefont {Mott}(1990)}]{mott1990}%
  \BibitemOpen
  \bibfield  {author} {\bibinfo {author} {\bibfnamefont {Neville}\ \bibnamefont
  {Mott}},\ }\href {\doibase https://doi.org/10.1201/b12795} {\emph {\bibinfo
  {title} {Metal-Insulator Transitions}}}\ (\bibinfo  {publisher} {CRC Press},\
  \bibinfo {address} {London},\ \bibinfo {year} {1990})\BibitemShut {NoStop}%
\bibitem [{\citenamefont {Luttinger}\ and\ \citenamefont {Ward}(1960)}]{LW}%
  \BibitemOpen
  \bibfield  {author} {\bibinfo {author} {\bibfnamefont {J.~M.}\ \bibnamefont
  {Luttinger}}\ and\ \bibinfo {author} {\bibfnamefont {J.~C.}\ \bibnamefont
  {Ward}},\ }\bibfield  {title} {\enquote {\bibinfo {title} {Ground-state
  energy of a many-fermion system. ii},}\ }\href {\doibase
  10.1103/PhysRev.118.1417} {\bibfield  {journal} {\bibinfo  {journal} {Phys.
  Rev.}\ }\textbf {\bibinfo {volume} {118}},\ \bibinfo {pages} {1417--1427}
  (\bibinfo {year} {1960})}\BibitemShut {NoStop}%
\bibitem [{\citenamefont {Oshikawa}(2000)}]{oshikawa}%
  \BibitemOpen
  \bibfield  {author} {\bibinfo {author} {\bibfnamefont {Masaki}\ \bibnamefont
  {Oshikawa}},\ }\bibfield  {title} {\enquote {\bibinfo {title} {Topological
  approach to luttinger's theorem and the fermi surface of a kondo lattice},}\
  }\href {\doibase 10.1103/PhysRevLett.84.3370} {\bibfield  {journal} {\bibinfo
   {journal} {Phys. Rev. Lett.}\ }\textbf {\bibinfo {volume} {84}},\ \bibinfo
  {pages} {3370--3373} (\bibinfo {year} {2000})}\BibitemShut {NoStop}%
\bibitem [{\citenamefont {Imada}\ \emph {et~al.}(1998)\citenamefont {Imada},
  \citenamefont {Fujimori},\ and\ \citenamefont {Tokura}}]{imada1998}%
  \BibitemOpen
  \bibfield  {author} {\bibinfo {author} {\bibfnamefont {Masatoshi}\
  \bibnamefont {Imada}}, \bibinfo {author} {\bibfnamefont {Atsushi}\
  \bibnamefont {Fujimori}}, \ and\ \bibinfo {author} {\bibfnamefont
  {Yoshinori}\ \bibnamefont {Tokura}},\ }\bibfield  {title} {\enquote {\bibinfo
  {title} {Metal-insulator transitions},}\ }\href {\doibase
  10.1103/RevModPhys.70.1039} {\bibfield  {journal} {\bibinfo  {journal} {Rev.
  Mod. Phys.}\ }\textbf {\bibinfo {volume} {70}},\ \bibinfo {pages}
  {1039--1263} (\bibinfo {year} {1998})}\BibitemShut {NoStop}%
\bibitem [{\citenamefont {Jamei}\ \emph {et~al.}(2005)\citenamefont {Jamei},
  \citenamefont {Kivelson},\ and\ \citenamefont
  {Spivak}}]{jamei_universal_2005}%
  \BibitemOpen
  \bibfield  {author} {\bibinfo {author} {\bibfnamefont {Reza}\ \bibnamefont
  {Jamei}}, \bibinfo {author} {\bibfnamefont {Steven}\ \bibnamefont
  {Kivelson}}, \ and\ \bibinfo {author} {\bibfnamefont {Boris}\ \bibnamefont
  {Spivak}},\ }\bibfield  {title} {\enquote {\bibinfo {title} {Universal
  {Aspects} of {Coulomb}-{Frustrated} {Phase} {Separation}},}\ }\href {\doibase
  10.1103/PhysRevLett.94.056805} {\bibfield  {journal} {\bibinfo  {journal}
  {Physical Review Letters}\ }\textbf {\bibinfo {volume} {94}},\ \bibinfo
  {pages} {056805} (\bibinfo {year} {2005})}\BibitemShut {NoStop}%
\bibitem [{\citenamefont {Camjayi}\ \emph {et~al.}(2008)\citenamefont
  {Camjayi}, \citenamefont {Haule}, \citenamefont {Dobrosavljevic},\ and\
  \citenamefont {Kotliar}}]{camjayi_coulomb_2008}%
  \BibitemOpen
  \bibfield  {author} {\bibinfo {author} {\bibfnamefont {A.}~\bibnamefont
  {Camjayi}}, \bibinfo {author} {\bibfnamefont {K.}~\bibnamefont {Haule}},
  \bibinfo {author} {\bibfnamefont {V.}~\bibnamefont {Dobrosavljevic}}, \ and\
  \bibinfo {author} {\bibfnamefont {G.}~\bibnamefont {Kotliar}},\ }\bibfield
  {title} {\enquote {\bibinfo {title} {Coulomb correlations and the
  {Wigner}–{Mott} transition},}\ }\href {\doibase 10.1038/nphys1106}
  {\bibfield  {journal} {\bibinfo  {journal} {Nature Physics}\ }\textbf
  {\bibinfo {volume} {4}},\ \bibinfo {pages} {932--935} (\bibinfo {year}
  {2008})}\BibitemShut {NoStop}%
\bibitem [{\citenamefont {Amaricci}\ \emph {et~al.}(2010)\citenamefont
  {Amaricci}, \citenamefont {Camjayi}, \citenamefont {Haule}, \citenamefont
  {Kotliar}, \citenamefont {Tanaskovic},\ and\ \citenamefont
  {Dobrosavljevic}}]{amaricci_extended_2010}%
  \BibitemOpen
  \bibfield  {author} {\bibinfo {author} {\bibfnamefont {A.}~\bibnamefont
  {Amaricci}}, \bibinfo {author} {\bibfnamefont {A.}~\bibnamefont {Camjayi}},
  \bibinfo {author} {\bibfnamefont {K.}~\bibnamefont {Haule}}, \bibinfo
  {author} {\bibfnamefont {G.}~\bibnamefont {Kotliar}}, \bibinfo {author}
  {\bibfnamefont {D.}~\bibnamefont {Tanaskovic}}, \ and\ \bibinfo {author}
  {\bibfnamefont {V.}~\bibnamefont {Dobrosavljevic}},\ }\bibfield  {title}
  {\enquote {\bibinfo {title} {Extended {Hubbard} model: {Charge} ordering and
  {Wigner}-{Mott} transition},}\ }\href {\doibase 10.1103/PhysRevB.82.155102}
  {\bibfield  {journal} {\bibinfo  {journal} {Physical Review B}\ }\textbf
  {\bibinfo {volume} {82}},\ \bibinfo {pages} {155102} (\bibinfo {year}
  {2010})}\BibitemShut {NoStop}%
\bibitem [{\citenamefont {Vu}\ and\ \citenamefont {Das~Sarma}(2020)}]{vu}%
  \BibitemOpen
  \bibfield  {author} {\bibinfo {author} {\bibfnamefont {DinhDuy}\ \bibnamefont
  {Vu}}\ and\ \bibinfo {author} {\bibfnamefont {S.}~\bibnamefont {Das~Sarma}},\
  }\bibfield  {title} {\enquote {\bibinfo {title} {One-dimensional few-electron
  effective wigner crystal in quantum and classical regimes},}\ }\href
  {\doibase 10.1103/PhysRevB.101.125113} {\bibfield  {journal} {\bibinfo
  {journal} {Phys. Rev. B}\ }\textbf {\bibinfo {volume} {101}},\ \bibinfo
  {pages} {125113} (\bibinfo {year} {2020})}\BibitemShut {NoStop}%
\bibitem [{\citenamefont {Musser}\ \emph {et~al.}(2022)\citenamefont {Musser},
  \citenamefont {Senthil},\ and\ \citenamefont
  {Chowdhury}}]{musser2022continuous}%
  \BibitemOpen
  \bibfield  {author} {\bibinfo {author} {\bibfnamefont {Seth}\ \bibnamefont
  {Musser}}, \bibinfo {author} {\bibfnamefont {T.}~\bibnamefont {Senthil}}, \
  and\ \bibinfo {author} {\bibfnamefont {Debanjan}\ \bibnamefont {Chowdhury}},\
  }\bibfield  {title} {\enquote {\bibinfo {title} {Theory of a continuous
  bandwidth-tuned wigner-mott transition},}\ }\href {\doibase
  10.1103/PhysRevB.106.155145} {\bibfield  {journal} {\bibinfo  {journal}
  {Phys. Rev. B}\ }\textbf {\bibinfo {volume} {106}},\ \bibinfo {pages}
  {155145} (\bibinfo {year} {2022})}\BibitemShut {NoStop}%
\bibitem [{\citenamefont {Xu}\ \emph {et~al.}(2022)\citenamefont {Xu},
  \citenamefont {Wu}, \citenamefont {Ye}, \citenamefont {Luo}, \citenamefont
  {Jian},\ and\ \citenamefont {Xu}}]{Xu22}%
  \BibitemOpen
  \bibfield  {author} {\bibinfo {author} {\bibfnamefont {Yichen}\ \bibnamefont
  {Xu}}, \bibinfo {author} {\bibfnamefont {Xiao-Chuan}\ \bibnamefont {Wu}},
  \bibinfo {author} {\bibfnamefont {Mengxing}\ \bibnamefont {Ye}}, \bibinfo
  {author} {\bibfnamefont {Zhu-Xi}\ \bibnamefont {Luo}}, \bibinfo {author}
  {\bibfnamefont {Chao-Ming}\ \bibnamefont {Jian}}, \ and\ \bibinfo {author}
  {\bibfnamefont {Cenke}\ \bibnamefont {Xu}},\ }\bibfield  {title} {\enquote
  {\bibinfo {title} {Interaction-driven metal-insulator transition with charge
  fractionalization},}\ }\href {\doibase 10.1103/PhysRevX.12.021067} {\bibfield
   {journal} {\bibinfo  {journal} {Phys. Rev. X}\ }\textbf {\bibinfo {volume}
  {12}},\ \bibinfo {pages} {021067} (\bibinfo {year} {2022})}\BibitemShut
  {NoStop}%
\bibitem [{\citenamefont {Cirac}\ \emph {et~al.}(2021)\citenamefont {Cirac},
  \citenamefont {P\'erez-Garc\'{\i}a}, \citenamefont {Schuch},\ and\
  \citenamefont {Verstraete}}]{RMP_MPS}%
  \BibitemOpen
  \bibfield  {author} {\bibinfo {author} {\bibfnamefont {J.~Ignacio}\
  \bibnamefont {Cirac}}, \bibinfo {author} {\bibfnamefont {David}\ \bibnamefont
  {P\'erez-Garc\'{\i}a}}, \bibinfo {author} {\bibfnamefont {Norbert}\
  \bibnamefont {Schuch}}, \ and\ \bibinfo {author} {\bibfnamefont {Frank}\
  \bibnamefont {Verstraete}},\ }\bibfield  {title} {\enquote {\bibinfo {title}
  {Matrix product states and projected entangled pair states: Concepts,
  symmetries, theorems},}\ }\href {\doibase 10.1103/RevModPhys.93.045003}
  {\bibfield  {journal} {\bibinfo  {journal} {Rev. Mod. Phys.}\ }\textbf
  {\bibinfo {volume} {93}},\ \bibinfo {pages} {045003} (\bibinfo {year}
  {2021})}\BibitemShut {NoStop}%
\bibitem [{\citenamefont {Schollwöck}(2011)}]{SCHOLLWOCKreview}%
  \BibitemOpen
  \bibfield  {author} {\bibinfo {author} {\bibfnamefont {Ulrich}\ \bibnamefont
  {Schollwöck}},\ }\bibfield  {title} {\enquote {\bibinfo {title} {The
  density-matrix renormalization group in the age of matrix product states},}\
  }\href {\doibase https://doi.org/10.1016/j.aop.2010.09.012} {\bibfield
  {journal} {\bibinfo  {journal} {Annals of Physics}\ }\textbf {\bibinfo
  {volume} {326}},\ \bibinfo {pages} {96--192} (\bibinfo {year} {2011})},\
  \bibinfo {note} {january 2011 Special Issue}\BibitemShut {NoStop}%
\bibitem [{\citenamefont {Luther}\ and\ \citenamefont
  {Emery}(1974)}]{luther1974}%
  \BibitemOpen
  \bibfield  {author} {\bibinfo {author} {\bibfnamefont {A.}~\bibnamefont
  {Luther}}\ and\ \bibinfo {author} {\bibfnamefont {V.~J.}\ \bibnamefont
  {Emery}},\ }\bibfield  {title} {\enquote {\bibinfo {title} {Backward
  scattering in the one-dimensional electron gas},}\ }\href {\doibase
  10.1103/PhysRevLett.33.589} {\bibfield  {journal} {\bibinfo  {journal} {Phys.
  Rev. Lett.}\ }\textbf {\bibinfo {volume} {33}},\ \bibinfo {pages} {589--592}
  (\bibinfo {year} {1974})}\BibitemShut {NoStop}%
\bibitem [{\citenamefont {{Carlson}}\ \emph {et~al.}(2002)\citenamefont
  {{Carlson}}, \citenamefont {{Emery}}, \citenamefont {{Kivelson}},\ and\
  \citenamefont {{Orgad}}}]{SKreview}%
  \BibitemOpen
  \bibfield  {author} {\bibinfo {author} {\bibfnamefont {E.~W.}\ \bibnamefont
  {{Carlson}}}, \bibinfo {author} {\bibfnamefont {V.~J.}\ \bibnamefont
  {{Emery}}}, \bibinfo {author} {\bibfnamefont {S.~A.}\ \bibnamefont
  {{Kivelson}}}, \ and\ \bibinfo {author} {\bibfnamefont {D.}~\bibnamefont
  {{Orgad}}},\ }\bibfield  {title} {\enquote {\bibinfo {title} {{Concepts in
  High Temperature Superconductivity}},}\ }\href {\doibase
  10.48550/arXiv.cond-mat/0206217} {\bibfield  {journal} {\bibinfo  {journal}
  {arXiv e-prints}\ ,\ \bibinfo {eid} {cond-mat/0206217}} (\bibinfo {year}
  {2002})},\ \Eprint {http://arxiv.org/abs/cond-mat/0206217}
  {arXiv:cond-mat/0206217 [cond-mat.supr-con]} \BibitemShut {NoStop}%
\bibitem [{\citenamefont {{Xia}}\ \emph {et~al.}(2024)\citenamefont {{Xia}},
  \citenamefont {{Han}}, \citenamefont {{Watanabe}}, \citenamefont
  {{Taniguchi}}, \citenamefont {{Shan}},\ and\ \citenamefont {{Mak}}}]{TMDSC}%
  \BibitemOpen
  \bibfield  {author} {\bibinfo {author} {\bibfnamefont {Yiyu}\ \bibnamefont
  {{Xia}}}, \bibinfo {author} {\bibfnamefont {Zhongdong}\ \bibnamefont
  {{Han}}}, \bibinfo {author} {\bibfnamefont {Kenji}\ \bibnamefont
  {{Watanabe}}}, \bibinfo {author} {\bibfnamefont {Takashi}\ \bibnamefont
  {{Taniguchi}}}, \bibinfo {author} {\bibfnamefont {Jie}\ \bibnamefont
  {{Shan}}}, \ and\ \bibinfo {author} {\bibfnamefont {Kin~Fai}\ \bibnamefont
  {{Mak}}},\ }\bibfield  {title} {\enquote {\bibinfo {title} {{Unconventional
  superconductivity in twisted bilayer WSe2}},}\ }\href {\doibase
  10.48550/arXiv.2405.14784} {\bibfield  {journal} {\bibinfo  {journal} {arXiv
  e-prints}\ ,\ \bibinfo {eid} {arXiv:2405.14784}} (\bibinfo {year} {2024})},\
  \Eprint {http://arxiv.org/abs/2405.14784} {arXiv:2405.14784
  [cond-mat.mes-hall]} \BibitemShut {NoStop}%
\bibitem [{\citenamefont {{Guo}}\ \emph {et~al.}(2024)\citenamefont {{Guo}},
  \citenamefont {{Pack}}, \citenamefont {{Swann}}, \citenamefont {{Holtzman}},
  \citenamefont {{Cothrine}}, \citenamefont {{Watanabe}}, \citenamefont
  {{Taniguchi}}, \citenamefont {{Mandrus}}, \citenamefont {{Barmak}},
  \citenamefont {{Hone}}, \citenamefont {{Millis}}, \citenamefont
  {{Pasupathy}},\ and\ \citenamefont {{Dean}}}]{Dean24}%
  \BibitemOpen
  \bibfield  {author} {\bibinfo {author} {\bibfnamefont {Yinjie}\ \bibnamefont
  {{Guo}}}, \bibinfo {author} {\bibfnamefont {Jordan}\ \bibnamefont {{Pack}}},
  \bibinfo {author} {\bibfnamefont {Joshua}\ \bibnamefont {{Swann}}}, \bibinfo
  {author} {\bibfnamefont {Luke}\ \bibnamefont {{Holtzman}}}, \bibinfo {author}
  {\bibfnamefont {Matthew}\ \bibnamefont {{Cothrine}}}, \bibinfo {author}
  {\bibfnamefont {Kenji}\ \bibnamefont {{Watanabe}}}, \bibinfo {author}
  {\bibfnamefont {Takashi}\ \bibnamefont {{Taniguchi}}}, \bibinfo {author}
  {\bibfnamefont {David}\ \bibnamefont {{Mandrus}}}, \bibinfo {author}
  {\bibfnamefont {Katayun}\ \bibnamefont {{Barmak}}}, \bibinfo {author}
  {\bibfnamefont {James}\ \bibnamefont {{Hone}}}, \bibinfo {author}
  {\bibfnamefont {Andrew~J.}\ \bibnamefont {{Millis}}}, \bibinfo {author}
  {\bibfnamefont {Abhay~N.}\ \bibnamefont {{Pasupathy}}}, \ and\ \bibinfo
  {author} {\bibfnamefont {Cory~R.}\ \bibnamefont {{Dean}}},\ }\bibfield
  {title} {\enquote {\bibinfo {title} {{Superconductivity in twisted bilayer
  WSe$_2$}},}\ }\href {\doibase 10.48550/arXiv.2406.03418} {\bibfield
  {journal} {\bibinfo  {journal} {arXiv e-prints}\ ,\ \bibinfo {eid}
  {arXiv:2406.03418}} (\bibinfo {year} {2024})},\ \Eprint
  {http://arxiv.org/abs/2406.03418} {arXiv:2406.03418 [cond-mat.mes-hall]}
  \BibitemShut {NoStop}%
\bibitem [{\citenamefont {Fabrizio}(1993)}]{fabrizio1993}%
  \BibitemOpen
  \bibfield  {author} {\bibinfo {author} {\bibfnamefont {M.}~\bibnamefont
  {Fabrizio}},\ }\bibfield  {title} {\enquote {\bibinfo {title} {Role of
  transverse hopping in a two-coupled-chains model},}\ }\href {\doibase
  10.1103/PhysRevB.48.15838} {\bibfield  {journal} {\bibinfo  {journal} {Phys.
  Rev. B}\ }\textbf {\bibinfo {volume} {48}},\ \bibinfo {pages} {15838--15860}
  (\bibinfo {year} {1993})}\BibitemShut {NoStop}%
\bibitem [{\citenamefont {Balents}\ and\ \citenamefont
  {Fisher}(1996)}]{balents1996}%
  \BibitemOpen
  \bibfield  {author} {\bibinfo {author} {\bibfnamefont {Leon}\ \bibnamefont
  {Balents}}\ and\ \bibinfo {author} {\bibfnamefont {Matthew P.~A.}\
  \bibnamefont {Fisher}},\ }\bibfield  {title} {\enquote {\bibinfo {title}
  {Weak-coupling phase diagram of the two-chain hubbard model},}\ }\href
  {\doibase 10.1103/PhysRevB.53.12133} {\bibfield  {journal} {\bibinfo
  {journal} {Phys. Rev. B}\ }\textbf {\bibinfo {volume} {53}},\ \bibinfo
  {pages} {12133--12141} (\bibinfo {year} {1996})}\BibitemShut {NoStop}%
\bibitem [{\citenamefont {Dagotto}\ and\ \citenamefont
  {Rice}(1996)}]{dagotto1996}%
  \BibitemOpen
  \bibfield  {author} {\bibinfo {author} {\bibfnamefont {Elbio}\ \bibnamefont
  {Dagotto}}\ and\ \bibinfo {author} {\bibfnamefont {T.~M.}\ \bibnamefont
  {Rice}},\ }\bibfield  {title} {\enquote {\bibinfo {title} {Surprises on the
  way from one- to two-dimensional quantum magnets: The ladder materials},}\
  }\href {\doibase 10.1126/science.271.5249.618} {\bibfield  {journal}
  {\bibinfo  {journal} {Science}\ }\textbf {\bibinfo {volume} {271}},\ \bibinfo
  {pages} {618--623} (\bibinfo {year} {1996})}\BibitemShut {NoStop}%
\bibitem [{\citenamefont {Kuroki}\ \emph {et~al.}(1997)\citenamefont {Kuroki},
  \citenamefont {Arita},\ and\ \citenamefont {Aoki}}]{kuroki}%
  \BibitemOpen
  \bibfield  {author} {\bibinfo {author} {\bibfnamefont {Kazuhiko}\
  \bibnamefont {Kuroki}}, \bibinfo {author} {\bibfnamefont {Ryotaro}\
  \bibnamefont {Arita}}, \ and\ \bibinfo {author} {\bibfnamefont {Hideo}\
  \bibnamefont {Aoki}},\ }\bibfield  {title} {\enquote {\bibinfo {title}
  {Numerical study of a superconductor-insulator transition in a half-filled
  hubbard chain with distant transfers},}\ }\href {\doibase
  10.1143/JPSJ.66.3371} {\bibfield  {journal} {\bibinfo  {journal} {Journal of
  the Physical Society of Japan}\ }\textbf {\bibinfo {volume} {66}},\ \bibinfo
  {pages} {3371--3374} (\bibinfo {year} {1997})}\BibitemShut {NoStop}%
\bibitem [{\citenamefont {Zhou}\ \emph {et~al.}(2023)\citenamefont {Zhou},
  \citenamefont {Ye}, \citenamefont {Luo}, \citenamefont {Zhao},\ and\
  \citenamefont {Chang}}]{zhou2023}%
  \BibitemOpen
  \bibfield  {author} {\bibinfo {author} {\bibfnamefont {Zongsheng}\
  \bibnamefont {Zhou}}, \bibinfo {author} {\bibfnamefont {Weinan}\ \bibnamefont
  {Ye}}, \bibinfo {author} {\bibfnamefont {Hong-Gang}\ \bibnamefont {Luo}},
  \bibinfo {author} {\bibfnamefont {Jize}\ \bibnamefont {Zhao}}, \ and\
  \bibinfo {author} {\bibfnamefont {Jun}\ \bibnamefont {Chang}},\ }\bibfield
  {title} {\enquote {\bibinfo {title} {Robust superconducting correlation
  against intersite interactions in the extended two-leg hubbard ladder},}\
  }\href {\doibase 10.1103/PhysRevB.108.195136} {\bibfield  {journal} {\bibinfo
   {journal} {Phys. Rev. B}\ }\textbf {\bibinfo {volume} {108}},\ \bibinfo
  {pages} {195136} (\bibinfo {year} {2023})}\BibitemShut {NoStop}%
\bibitem [{\citenamefont {Musser}\ and\ \citenamefont
  {Senthil}(2022)}]{musser2022}%
  \BibitemOpen
  \bibfield  {author} {\bibinfo {author} {\bibfnamefont {Seth}\ \bibnamefont
  {Musser}}\ and\ \bibinfo {author} {\bibfnamefont {T.}~\bibnamefont
  {Senthil}},\ }\bibfield  {title} {\enquote {\bibinfo {title} {Metal to
  wigner-mott insulator transition in two-leg ladders},}\ }\href {\doibase
  10.1103/PhysRevB.106.235148} {\bibfield  {journal} {\bibinfo  {journal}
  {Phys. Rev. B}\ }\textbf {\bibinfo {volume} {106}},\ \bibinfo {pages}
  {235148} (\bibinfo {year} {2022})}\BibitemShut {NoStop}%
\bibitem [{\citenamefont {Li}\ \emph {et~al.}(2021{\natexlab{a}})\citenamefont
  {Li}, \citenamefont {Jiang}, \citenamefont {Li}, \citenamefont {Zhang},
  \citenamefont {Kang}, \citenamefont {Zhu}, \citenamefont {Watanabe},
  \citenamefont {Taniguchi}, \citenamefont {Chowdhury}, \citenamefont {Fu}
  \emph {et~al.}}]{li2021continuous}%
  \BibitemOpen
  \bibfield  {author} {\bibinfo {author} {\bibfnamefont {Tingxin}\ \bibnamefont
  {Li}}, \bibinfo {author} {\bibfnamefont {Shengwei}\ \bibnamefont {Jiang}},
  \bibinfo {author} {\bibfnamefont {Lizhong}\ \bibnamefont {Li}}, \bibinfo
  {author} {\bibfnamefont {Yang}\ \bibnamefont {Zhang}}, \bibinfo {author}
  {\bibfnamefont {Kaifei}\ \bibnamefont {Kang}}, \bibinfo {author}
  {\bibfnamefont {Jiacheng}\ \bibnamefont {Zhu}}, \bibinfo {author}
  {\bibfnamefont {Kenji}\ \bibnamefont {Watanabe}}, \bibinfo {author}
  {\bibfnamefont {Takashi}\ \bibnamefont {Taniguchi}}, \bibinfo {author}
  {\bibfnamefont {Debanjan}\ \bibnamefont {Chowdhury}}, \bibinfo {author}
  {\bibfnamefont {Liang}\ \bibnamefont {Fu}},  \emph {et~al.},\ }\bibfield
  {title} {\enquote {\bibinfo {title} {Continuous mott transition in
  semiconductor moir{\'e} superlattices},}\ }\href {\doibase
  https://doi.org/10.1038/s41586-021-03853-0} {\bibfield  {journal} {\bibinfo
  {journal} {Nature}\ }\textbf {\bibinfo {volume} {597}},\ \bibinfo {pages}
  {350--354} (\bibinfo {year} {2021}{\natexlab{a}})}\BibitemShut {NoStop}%
\bibitem [{\citenamefont {Ghiotto}\ \emph {et~al.}(2021)\citenamefont
  {Ghiotto}, \citenamefont {Shih}, \citenamefont {Pereira}, \citenamefont
  {Rhodes}, \citenamefont {Kim}, \citenamefont {Zang}, \citenamefont {Millis},
  \citenamefont {Watanabe}, \citenamefont {Taniguchi}, \citenamefont {Hone}
  \emph {et~al.}}]{ghiotto2021quantum}%
  \BibitemOpen
  \bibfield  {author} {\bibinfo {author} {\bibfnamefont {Augusto}\ \bibnamefont
  {Ghiotto}}, \bibinfo {author} {\bibfnamefont {En-Min}\ \bibnamefont {Shih}},
  \bibinfo {author} {\bibfnamefont {Giancarlo~SSG}\ \bibnamefont {Pereira}},
  \bibinfo {author} {\bibfnamefont {Daniel~A}\ \bibnamefont {Rhodes}}, \bibinfo
  {author} {\bibfnamefont {Bumho}\ \bibnamefont {Kim}}, \bibinfo {author}
  {\bibfnamefont {Jiawei}\ \bibnamefont {Zang}}, \bibinfo {author}
  {\bibfnamefont {Andrew~J}\ \bibnamefont {Millis}}, \bibinfo {author}
  {\bibfnamefont {Kenji}\ \bibnamefont {Watanabe}}, \bibinfo {author}
  {\bibfnamefont {Takashi}\ \bibnamefont {Taniguchi}}, \bibinfo {author}
  {\bibfnamefont {James~C}\ \bibnamefont {Hone}},  \emph {et~al.},\ }\bibfield
  {title} {\enquote {\bibinfo {title} {Quantum criticality in twisted
  transition metal dichalcogenides},}\ }\href {\doibase
  https://doi.org/10.1038/s41586-021-03815-6} {\bibfield  {journal} {\bibinfo
  {journal} {Nature}\ }\textbf {\bibinfo {volume} {597}},\ \bibinfo {pages}
  {345--349} (\bibinfo {year} {2021})}\BibitemShut {NoStop}%
\bibitem [{\citenamefont {Xu}\ \emph {et~al.}(2020)\citenamefont {Xu},
  \citenamefont {Liu}, \citenamefont {Rhodes}, \citenamefont {Watanabe},
  \citenamefont {Taniguchi}, \citenamefont {Hone}, \citenamefont {Elser},
  \citenamefont {Mak},\ and\ \citenamefont {Shan}}]{xu2020correlated}%
  \BibitemOpen
  \bibfield  {author} {\bibinfo {author} {\bibfnamefont {Yang}\ \bibnamefont
  {Xu}}, \bibinfo {author} {\bibfnamefont {Song}\ \bibnamefont {Liu}}, \bibinfo
  {author} {\bibfnamefont {Daniel~A}\ \bibnamefont {Rhodes}}, \bibinfo {author}
  {\bibfnamefont {Kenji}\ \bibnamefont {Watanabe}}, \bibinfo {author}
  {\bibfnamefont {Takashi}\ \bibnamefont {Taniguchi}}, \bibinfo {author}
  {\bibfnamefont {James}\ \bibnamefont {Hone}}, \bibinfo {author}
  {\bibfnamefont {Veit}\ \bibnamefont {Elser}}, \bibinfo {author}
  {\bibfnamefont {Kin~Fai}\ \bibnamefont {Mak}}, \ and\ \bibinfo {author}
  {\bibfnamefont {Jie}\ \bibnamefont {Shan}},\ }\bibfield  {title} {\enquote
  {\bibinfo {title} {Correlated insulating states at fractional fillings of
  moir{\'e} superlattices},}\ }\href {\doibase
  https://doi.org/10.1038/s41586-020-2868-6} {\bibfield  {journal} {\bibinfo
  {journal} {Nature}\ }\textbf {\bibinfo {volume} {587}},\ \bibinfo {pages}
  {214--218} (\bibinfo {year} {2020})}\BibitemShut {NoStop}%
\bibitem [{\citenamefont {Regan}\ \emph {et~al.}(2020)\citenamefont {Regan},
  \citenamefont {Wang}, \citenamefont {Jin}, \citenamefont {Bakti~Utama},
  \citenamefont {Gao}, \citenamefont {Wei}, \citenamefont {Zhao}, \citenamefont
  {Zhao}, \citenamefont {Zhang}, \citenamefont {Yumigeta}, \citenamefont
  {Blei}, \citenamefont {Carlström}, \citenamefont {Watanabe}, \citenamefont
  {Taniguchi}, \citenamefont {Tongay}, \citenamefont {Crommie}, \citenamefont
  {Zettl},\ and\ \citenamefont {Wang}}]{regan_mott_2020}%
  \BibitemOpen
  \bibfield  {author} {\bibinfo {author} {\bibfnamefont {Emma~C.}\ \bibnamefont
  {Regan}}, \bibinfo {author} {\bibfnamefont {Danqing}\ \bibnamefont {Wang}},
  \bibinfo {author} {\bibfnamefont {Chenhao}\ \bibnamefont {Jin}}, \bibinfo
  {author} {\bibfnamefont {M.~Iqbal}\ \bibnamefont {Bakti~Utama}}, \bibinfo
  {author} {\bibfnamefont {Beini}\ \bibnamefont {Gao}}, \bibinfo {author}
  {\bibfnamefont {Xin}\ \bibnamefont {Wei}}, \bibinfo {author} {\bibfnamefont
  {Sihan}\ \bibnamefont {Zhao}}, \bibinfo {author} {\bibfnamefont {Wenyu}\
  \bibnamefont {Zhao}}, \bibinfo {author} {\bibfnamefont {Zuocheng}\
  \bibnamefont {Zhang}}, \bibinfo {author} {\bibfnamefont {Kentaro}\
  \bibnamefont {Yumigeta}}, \bibinfo {author} {\bibfnamefont {Mark}\
  \bibnamefont {Blei}}, \bibinfo {author} {\bibfnamefont {Johan~D.}\
  \bibnamefont {Carlström}}, \bibinfo {author} {\bibfnamefont {Kenji}\
  \bibnamefont {Watanabe}}, \bibinfo {author} {\bibfnamefont {Takashi}\
  \bibnamefont {Taniguchi}}, \bibinfo {author} {\bibfnamefont {Sefaattin}\
  \bibnamefont {Tongay}}, \bibinfo {author} {\bibfnamefont {Michael}\
  \bibnamefont {Crommie}}, \bibinfo {author} {\bibfnamefont {Alex}\
  \bibnamefont {Zettl}}, \ and\ \bibinfo {author} {\bibfnamefont {Feng}\
  \bibnamefont {Wang}},\ }\bibfield  {title} {\enquote {\bibinfo {title} {Mott
  and generalized {Wigner} crystal states in {WSe2}/{WS2} moiré
  superlattices},}\ }\href {\doibase 10.1038/s41586-020-2092-4} {\bibfield
  {journal} {\bibinfo  {journal} {Nature}\ }\textbf {\bibinfo {volume} {579}},\
  \bibinfo {pages} {359--363} (\bibinfo {year} {2020})}\BibitemShut {NoStop}%
\bibitem [{\citenamefont {Huang}\ \emph {et~al.}(2021)\citenamefont {Huang},
  \citenamefont {Wang}, \citenamefont {Miao}, \citenamefont {Wang},
  \citenamefont {Li}, \citenamefont {Lian}, \citenamefont {Taniguchi},
  \citenamefont {Watanabe}, \citenamefont {Okamoto}, \citenamefont {Xiao},
  \citenamefont {Shi},\ and\ \citenamefont {Cui}}]{huang_correlated_2021}%
  \BibitemOpen
  \bibfield  {author} {\bibinfo {author} {\bibfnamefont {Xiong}\ \bibnamefont
  {Huang}}, \bibinfo {author} {\bibfnamefont {Tianmeng}\ \bibnamefont {Wang}},
  \bibinfo {author} {\bibfnamefont {Shengnan}\ \bibnamefont {Miao}}, \bibinfo
  {author} {\bibfnamefont {Chong}\ \bibnamefont {Wang}}, \bibinfo {author}
  {\bibfnamefont {Zhipeng}\ \bibnamefont {Li}}, \bibinfo {author}
  {\bibfnamefont {Zhen}\ \bibnamefont {Lian}}, \bibinfo {author} {\bibfnamefont
  {Takashi}\ \bibnamefont {Taniguchi}}, \bibinfo {author} {\bibfnamefont
  {Kenji}\ \bibnamefont {Watanabe}}, \bibinfo {author} {\bibfnamefont
  {Satoshi}\ \bibnamefont {Okamoto}}, \bibinfo {author} {\bibfnamefont
  {Di}~\bibnamefont {Xiao}}, \bibinfo {author} {\bibfnamefont {Su-Fei}\
  \bibnamefont {Shi}}, \ and\ \bibinfo {author} {\bibfnamefont {Yong-Tao}\
  \bibnamefont {Cui}},\ }\bibfield  {title} {\enquote {\bibinfo {title}
  {Correlated insulating states at fractional fillings of the {WS2}/{WSe2}
  moiré lattice},}\ }\href {\doibase 10.1038/s41567-021-01171-w} {\bibfield
  {journal} {\bibinfo  {journal} {Nature Physics}\ }\textbf {\bibinfo {volume}
  {17}},\ \bibinfo {pages} {715--719} (\bibinfo {year} {2021})}\BibitemShut
  {NoStop}%
\bibitem [{\citenamefont {Jiang}\ \emph {et~al.}(1990)\citenamefont {Jiang},
  \citenamefont {Willett}, \citenamefont {Stormer}, \citenamefont {Tsui},
  \citenamefont {Pfeiffer},\ and\ \citenamefont {West}}]{QHWC1}%
  \BibitemOpen
  \bibfield  {author} {\bibinfo {author} {\bibfnamefont {H.~W.}\ \bibnamefont
  {Jiang}}, \bibinfo {author} {\bibfnamefont {R.~L.}\ \bibnamefont {Willett}},
  \bibinfo {author} {\bibfnamefont {H.~L.}\ \bibnamefont {Stormer}}, \bibinfo
  {author} {\bibfnamefont {D.~C.}\ \bibnamefont {Tsui}}, \bibinfo {author}
  {\bibfnamefont {L.~N.}\ \bibnamefont {Pfeiffer}}, \ and\ \bibinfo {author}
  {\bibfnamefont {K.~W.}\ \bibnamefont {West}},\ }\bibfield  {title} {\enquote
  {\bibinfo {title} {Quantum liquid versus electron solid around
  \ensuremath{\nu}=1/5 landau-level filling},}\ }\href {\doibase
  10.1103/PhysRevLett.65.633} {\bibfield  {journal} {\bibinfo  {journal} {Phys.
  Rev. Lett.}\ }\textbf {\bibinfo {volume} {65}},\ \bibinfo {pages} {633--636}
  (\bibinfo {year} {1990})}\BibitemShut {NoStop}%
\bibitem [{\citenamefont {Yoon}\ \emph {et~al.}(1999)\citenamefont {Yoon},
  \citenamefont {Li}, \citenamefont {Shahar}, \citenamefont {Tsui},\ and\
  \citenamefont {Shayegan}}]{QHWC2}%
  \BibitemOpen
  \bibfield  {author} {\bibinfo {author} {\bibfnamefont {Jongsoo}\ \bibnamefont
  {Yoon}}, \bibinfo {author} {\bibfnamefont {C.~C.}\ \bibnamefont {Li}},
  \bibinfo {author} {\bibfnamefont {D.}~\bibnamefont {Shahar}}, \bibinfo
  {author} {\bibfnamefont {D.~C.}\ \bibnamefont {Tsui}}, \ and\ \bibinfo
  {author} {\bibfnamefont {M.}~\bibnamefont {Shayegan}},\ }\bibfield  {title}
  {\enquote {\bibinfo {title} {Wigner crystallization and metal-insulator
  transition of two-dimensional holes in gaas at
  $\mathit{B}\phantom{\rule{0ex}{0ex}}=\phantom{\rule{0ex}{0ex}}0$},}\ }\href
  {\doibase 10.1103/PhysRevLett.82.1744} {\bibfield  {journal} {\bibinfo
  {journal} {Phys. Rev. Lett.}\ }\textbf {\bibinfo {volume} {82}},\ \bibinfo
  {pages} {1744--1747} (\bibinfo {year} {1999})}\BibitemShut {NoStop}%
\bibitem [{\citenamefont {Hossain}\ \emph {et~al.}(2020)\citenamefont
  {Hossain}, \citenamefont {Ma}, \citenamefont {Rosales}, \citenamefont
  {Chung}, \citenamefont {Pfeiffer}, \citenamefont {West}, \citenamefont
  {Baldwin},\ and\ \citenamefont {Shayegan}}]{QHWC3}%
  \BibitemOpen
  \bibfield  {author} {\bibinfo {author} {\bibfnamefont {Md~S}\ \bibnamefont
  {Hossain}}, \bibinfo {author} {\bibfnamefont {MK}~\bibnamefont {Ma}},
  \bibinfo {author} {\bibfnamefont {KA}~\bibnamefont {Rosales}}, \bibinfo
  {author} {\bibfnamefont {YJ}~\bibnamefont {Chung}}, \bibinfo {author}
  {\bibfnamefont {LN}~\bibnamefont {Pfeiffer}}, \bibinfo {author}
  {\bibfnamefont {KW}~\bibnamefont {West}}, \bibinfo {author} {\bibfnamefont
  {KW}~\bibnamefont {Baldwin}}, \ and\ \bibinfo {author} {\bibfnamefont
  {Mansour}\ \bibnamefont {Shayegan}},\ }\bibfield  {title} {\enquote {\bibinfo
  {title} {Observation of spontaneous ferromagnetism in a two-dimensional
  electron system},}\ }\href@noop {} {\bibfield  {journal} {\bibinfo  {journal}
  {Proceedings of the National Academy of Sciences}\ }\textbf {\bibinfo
  {volume} {117}},\ \bibinfo {pages} {32244--32250} (\bibinfo {year}
  {2020})}\BibitemShut {NoStop}%
\bibitem [{\citenamefont {Smole{\'{n}}ski}\ \emph {et~al.}(2021)\citenamefont
  {Smole{\'{n}}ski}, \citenamefont {Dolgirev}, \citenamefont {Kuhlenkamp},
  \citenamefont {Popert}, \citenamefont {Shimazaki}, \citenamefont {Back},
  \citenamefont {Lu}, \citenamefont {Kroner}, \citenamefont {Watanabe},
  \citenamefont {Taniguchi}, \citenamefont {Esterlis}, \citenamefont {Demler},\
  and\ \citenamefont {Imamo{\u{g}}lu}}]{TMDWC1}%
  \BibitemOpen
  \bibfield  {author} {\bibinfo {author} {\bibfnamefont {Tomasz}\ \bibnamefont
  {Smole{\'{n}}ski}}, \bibinfo {author} {\bibfnamefont {Pavel~E.}\ \bibnamefont
  {Dolgirev}}, \bibinfo {author} {\bibfnamefont {Clemens}\ \bibnamefont
  {Kuhlenkamp}}, \bibinfo {author} {\bibfnamefont {Alexander}\ \bibnamefont
  {Popert}}, \bibinfo {author} {\bibfnamefont {Yuya}\ \bibnamefont
  {Shimazaki}}, \bibinfo {author} {\bibfnamefont {Patrick}\ \bibnamefont
  {Back}}, \bibinfo {author} {\bibfnamefont {Xiaobo}\ \bibnamefont {Lu}},
  \bibinfo {author} {\bibfnamefont {Martin}\ \bibnamefont {Kroner}}, \bibinfo
  {author} {\bibfnamefont {Kenji}\ \bibnamefont {Watanabe}}, \bibinfo {author}
  {\bibfnamefont {Takashi}\ \bibnamefont {Taniguchi}}, \bibinfo {author}
  {\bibfnamefont {Ilya}\ \bibnamefont {Esterlis}}, \bibinfo {author}
  {\bibfnamefont {Eugene}\ \bibnamefont {Demler}}, \ and\ \bibinfo {author}
  {\bibfnamefont {Ata{\c{c}}}\ \bibnamefont {Imamo{\u{g}}lu}},\ }\bibfield
  {title} {\enquote {\bibinfo {title} {Signatures of wigner crystal of
  electrons in a monolayer semiconductor},}\ }\href {\doibase
  10.1038/s41586-021-03590-4} {\bibfield  {journal} {\bibinfo  {journal}
  {Nature}\ }\textbf {\bibinfo {volume} {595}},\ \bibinfo {pages} {53--57}
  (\bibinfo {year} {2021})}\BibitemShut {NoStop}%
\bibitem [{\citenamefont {Zhou}\ \emph {et~al.}(2021)\citenamefont {Zhou},
  \citenamefont {Sung}, \citenamefont {Brutschea}, \citenamefont {Esterlis},
  \citenamefont {Wang}, \citenamefont {Scuri}, \citenamefont {Gelly},
  \citenamefont {Heo}, \citenamefont {Taniguchi}, \citenamefont {Watanabe},
  \citenamefont {Zar{\'a}nd}, \citenamefont {Lukin}, \citenamefont {Kim},
  \citenamefont {Demler},\ and\ \citenamefont {Park}}]{TMDWC2}%
  \BibitemOpen
  \bibfield  {author} {\bibinfo {author} {\bibfnamefont {You}\ \bibnamefont
  {Zhou}}, \bibinfo {author} {\bibfnamefont {Jiho}\ \bibnamefont {Sung}},
  \bibinfo {author} {\bibfnamefont {Elise}\ \bibnamefont {Brutschea}}, \bibinfo
  {author} {\bibfnamefont {Ilya}\ \bibnamefont {Esterlis}}, \bibinfo {author}
  {\bibfnamefont {Yao}\ \bibnamefont {Wang}}, \bibinfo {author} {\bibfnamefont
  {Giovanni}\ \bibnamefont {Scuri}}, \bibinfo {author} {\bibfnamefont
  {Ryan~J.}\ \bibnamefont {Gelly}}, \bibinfo {author} {\bibfnamefont {Hoseok}\
  \bibnamefont {Heo}}, \bibinfo {author} {\bibfnamefont {Takashi}\ \bibnamefont
  {Taniguchi}}, \bibinfo {author} {\bibfnamefont {Kenji}\ \bibnamefont
  {Watanabe}}, \bibinfo {author} {\bibfnamefont {Gergely}\ \bibnamefont
  {Zar{\'a}nd}}, \bibinfo {author} {\bibfnamefont {Mikhail~D.}\ \bibnamefont
  {Lukin}}, \bibinfo {author} {\bibfnamefont {Philip}\ \bibnamefont {Kim}},
  \bibinfo {author} {\bibfnamefont {Eugene}\ \bibnamefont {Demler}}, \ and\
  \bibinfo {author} {\bibfnamefont {Hongkun}\ \bibnamefont {Park}},\ }\bibfield
   {title} {\enquote {\bibinfo {title} {Bilayer wigner crystals in a transition
  metal dichalcogenide heterostructure},}\ }\href {\doibase
  10.1038/s41586-021-03560-w} {\bibfield  {journal} {\bibinfo  {journal}
  {Nature}\ }\textbf {\bibinfo {volume} {595}},\ \bibinfo {pages} {48--52}
  (\bibinfo {year} {2021})}\BibitemShut {NoStop}%
\bibitem [{\citenamefont {Li}\ \emph {et~al.}(2021{\natexlab{b}})\citenamefont
  {Li}, \citenamefont {Li}, \citenamefont {Regan}, \citenamefont {Wang},
  \citenamefont {Zhao}, \citenamefont {Kahn}, \citenamefont {Yumigeta},
  \citenamefont {Blei}, \citenamefont {Taniguchi}, \citenamefont {Watanabe},
  \citenamefont {Tongay}, \citenamefont {Zettl}, \citenamefont {Crommie},\ and\
  \citenamefont {Wang}}]{li_imaging_2021}%
  \BibitemOpen
  \bibfield  {author} {\bibinfo {author} {\bibfnamefont {Hongyuan}\
  \bibnamefont {Li}}, \bibinfo {author} {\bibfnamefont {Shaowei}\ \bibnamefont
  {Li}}, \bibinfo {author} {\bibfnamefont {Emma~C.}\ \bibnamefont {Regan}},
  \bibinfo {author} {\bibfnamefont {Danqing}\ \bibnamefont {Wang}}, \bibinfo
  {author} {\bibfnamefont {Wenyu}\ \bibnamefont {Zhao}}, \bibinfo {author}
  {\bibfnamefont {Salman}\ \bibnamefont {Kahn}}, \bibinfo {author}
  {\bibfnamefont {Kentaro}\ \bibnamefont {Yumigeta}}, \bibinfo {author}
  {\bibfnamefont {Mark}\ \bibnamefont {Blei}}, \bibinfo {author} {\bibfnamefont
  {Takashi}\ \bibnamefont {Taniguchi}}, \bibinfo {author} {\bibfnamefont
  {Kenji}\ \bibnamefont {Watanabe}}, \bibinfo {author} {\bibfnamefont
  {Sefaattin}\ \bibnamefont {Tongay}}, \bibinfo {author} {\bibfnamefont {Alex}\
  \bibnamefont {Zettl}}, \bibinfo {author} {\bibfnamefont {Michael~F.}\
  \bibnamefont {Crommie}}, \ and\ \bibinfo {author} {\bibfnamefont {Feng}\
  \bibnamefont {Wang}},\ }\bibfield  {title} {\enquote {\bibinfo {title}
  {Imaging two-dimensional generalized {Wigner} crystals},}\ }\href {\doibase
  10.1038/s41586-021-03874-9} {\bibfield  {journal} {\bibinfo  {journal}
  {Nature}\ }\textbf {\bibinfo {volume} {597}},\ \bibinfo {pages} {650--654}
  (\bibinfo {year} {2021}{\natexlab{b}})}\BibitemShut {NoStop}%
\bibitem [{\citenamefont {Roger}(1984)}]{roger}%
  \BibitemOpen
  \bibfield  {author} {\bibinfo {author} {\bibfnamefont {M.}~\bibnamefont
  {Roger}},\ }\bibfield  {title} {\enquote {\bibinfo {title} {Multiple exchange
  in $^{3}\mathrm{He}$ and in the wigner solid},}\ }\href {\doibase
  10.1103/PhysRevB.30.6432} {\bibfield  {journal} {\bibinfo  {journal} {Phys.
  Rev. B}\ }\textbf {\bibinfo {volume} {30}},\ \bibinfo {pages} {6432--6457}
  (\bibinfo {year} {1984})}\BibitemShut {NoStop}%
\bibitem [{\citenamefont {Chakravarty}\ \emph {et~al.}(1999)\citenamefont
  {Chakravarty}, \citenamefont {Kivelson}, \citenamefont {Nayak},\ and\
  \citenamefont {Voelker}}]{SK99}%
  \BibitemOpen
  \bibfield  {author} {\bibinfo {author} {\bibfnamefont {Sudip}\ \bibnamefont
  {Chakravarty}}, \bibinfo {author} {\bibfnamefont {Steven}\ \bibnamefont
  {Kivelson}}, \bibinfo {author} {\bibfnamefont {Cheta}\ \bibnamefont {Nayak}},
  \ and\ \bibinfo {author} {\bibfnamefont {Klaus}\ \bibnamefont {Voelker}},\
  }\bibfield  {title} {\enquote {\bibinfo {title} {Wigner glass, spin liquids
  and the metal-insulator transition},}\ }\href {\doibase
  10.1080/13642819908214845} {\bibfield  {journal} {\bibinfo  {journal}
  {Philosophical Magazine B}\ }\textbf {\bibinfo {volume} {79}},\ \bibinfo
  {pages} {859--868} (\bibinfo {year} {1999})},\ \Eprint
  {http://arxiv.org/abs/https://doi.org/10.1080/13642819908214845}
  {https://doi.org/10.1080/13642819908214845} \BibitemShut {NoStop}%
\bibitem [{\citenamefont {Spivak}\ and\ \citenamefont
  {Kivelson}(2006)}]{spivak}%
  \BibitemOpen
  \bibfield  {author} {\bibinfo {author} {\bibfnamefont {Boris}\ \bibnamefont
  {Spivak}}\ and\ \bibinfo {author} {\bibfnamefont {Steven~A.}\ \bibnamefont
  {Kivelson}},\ }\bibfield  {title} {\enquote {\bibinfo {title} {Transport in
  two dimensional electronic micro-emulsions},}\ }\href {\doibase
  https://doi.org/10.1016/j.aop.2005.12.002} {\bibfield  {journal} {\bibinfo
  {journal} {Annals of Physics}\ }\textbf {\bibinfo {volume} {321}},\ \bibinfo
  {pages} {2071--2115} (\bibinfo {year} {2006})}\BibitemShut {NoStop}%
\bibitem [{\citenamefont {Padhi}\ \emph {et~al.}(2021)\citenamefont {Padhi},
  \citenamefont {Chitra},\ and\ \citenamefont {Phillips}}]{PP21}%
  \BibitemOpen
  \bibfield  {author} {\bibinfo {author} {\bibfnamefont {Bikash}\ \bibnamefont
  {Padhi}}, \bibinfo {author} {\bibfnamefont {R.}~\bibnamefont {Chitra}}, \
  and\ \bibinfo {author} {\bibfnamefont {Philip~W.}\ \bibnamefont {Phillips}},\
  }\bibfield  {title} {\enquote {\bibinfo {title} {Generalized wigner
  crystallization in moir\'e materials},}\ }\href {\doibase
  10.1103/PhysRevB.103.125146} {\bibfield  {journal} {\bibinfo  {journal}
  {Phys. Rev. B}\ }\textbf {\bibinfo {volume} {103}},\ \bibinfo {pages}
  {125146} (\bibinfo {year} {2021})}\BibitemShut {NoStop}%
\bibitem [{\citenamefont {Zhang}\ \emph {et~al.}(2021)\citenamefont {Zhang},
  \citenamefont {Liu},\ and\ \citenamefont {Fu}}]{LF21}%
  \BibitemOpen
  \bibfield  {author} {\bibinfo {author} {\bibfnamefont {Yang}\ \bibnamefont
  {Zhang}}, \bibinfo {author} {\bibfnamefont {Tongtong}\ \bibnamefont {Liu}}, \
  and\ \bibinfo {author} {\bibfnamefont {Liang}\ \bibnamefont {Fu}},\
  }\bibfield  {title} {\enquote {\bibinfo {title} {Electronic structures,
  charge transfer, and charge order in twisted transition metal dichalcogenide
  bilayers},}\ }\href {\doibase 10.1103/PhysRevB.103.155142} {\bibfield
  {journal} {\bibinfo  {journal} {Phys. Rev. B}\ }\textbf {\bibinfo {volume}
  {103}},\ \bibinfo {pages} {155142} (\bibinfo {year} {2021})}\BibitemShut
  {NoStop}%
\bibitem [{\citenamefont {Pan}\ \emph {et~al.}(2020)\citenamefont {Pan},
  \citenamefont {Wu},\ and\ \citenamefont {Das~Sarma}}]{SDS20}%
  \BibitemOpen
  \bibfield  {author} {\bibinfo {author} {\bibfnamefont {Haining}\ \bibnamefont
  {Pan}}, \bibinfo {author} {\bibfnamefont {Fengcheng}\ \bibnamefont {Wu}}, \
  and\ \bibinfo {author} {\bibfnamefont {Sankar}\ \bibnamefont {Das~Sarma}},\
  }\bibfield  {title} {\enquote {\bibinfo {title} {Quantum phase diagram of a
  moir\'e-hubbard model},}\ }\href {\doibase 10.1103/PhysRevB.102.201104}
  {\bibfield  {journal} {\bibinfo  {journal} {Phys. Rev. B}\ }\textbf {\bibinfo
  {volume} {102}},\ \bibinfo {pages} {201104} (\bibinfo {year}
  {2020})}\BibitemShut {NoStop}%
\bibitem [{\citenamefont {Pan}\ and\ \citenamefont
  {Das~Sarma}(2021)}]{pan_interaction-driven_2021}%
  \BibitemOpen
  \bibfield  {author} {\bibinfo {author} {\bibfnamefont {Haining}\ \bibnamefont
  {Pan}}\ and\ \bibinfo {author} {\bibfnamefont {Sankar}\ \bibnamefont
  {Das~Sarma}},\ }\bibfield  {title} {\enquote {\bibinfo {title}
  {Interaction-{Driven} {Filling}-{Induced} {Metal}-{Insulator} {Transitions}
  in {2D} {Moiré} {Lattices}},}\ }\href {\doibase
  10.1103/PhysRevLett.127.096802} {\bibfield  {journal} {\bibinfo  {journal}
  {Physical Review Letters}\ }\textbf {\bibinfo {volume} {127}},\ \bibinfo
  {pages} {096802} (\bibinfo {year} {2021})}\BibitemShut {NoStop}%
\bibitem [{\citenamefont {Pan}\ and\ \citenamefont {Das~Sarma}(2022)}]{SDS22}%
  \BibitemOpen
  \bibfield  {author} {\bibinfo {author} {\bibfnamefont {Haining}\ \bibnamefont
  {Pan}}\ and\ \bibinfo {author} {\bibfnamefont {Sankar}\ \bibnamefont
  {Das~Sarma}},\ }\bibfield  {title} {\enquote {\bibinfo {title} {Interaction
  range and temperature dependence of symmetry breaking in strongly correlated
  two-dimensional moir\'e transition metal dichalcogenide bilayers},}\ }\href
  {\doibase 10.1103/PhysRevB.105.041109} {\bibfield  {journal} {\bibinfo
  {journal} {Phys. Rev. B}\ }\textbf {\bibinfo {volume} {105}},\ \bibinfo
  {pages} {041109} (\bibinfo {year} {2022})}\BibitemShut {NoStop}%
\bibitem [{\citenamefont {{Matty}}\ and\ \citenamefont {{Kim}}(2022)}]{matty}%
  \BibitemOpen
  \bibfield  {author} {\bibinfo {author} {\bibfnamefont {Michael}\ \bibnamefont
  {{Matty}}}\ and\ \bibinfo {author} {\bibfnamefont {Eun-Ah}\ \bibnamefont
  {{Kim}}},\ }\bibfield  {title} {\enquote {\bibinfo {title} {{Melting of
  generalized Wigner crystals in transition metal dichalcogenide heterobilayer
  Moir{\'e} systems}},}\ }\href {\doibase 10.1038/s41467-022-34683-x}
  {\bibfield  {journal} {\bibinfo  {journal} {Nature Communications}\ }\textbf
  {\bibinfo {volume} {13}},\ \bibinfo {eid} {7098} (\bibinfo {year}
  {2022})}\BibitemShut {NoStop}%
\bibitem [{\citenamefont {Zhou}\ \emph {et~al.}(2022)\citenamefont {Zhou},
  \citenamefont {Sheng},\ and\ \citenamefont {Kim}}]{Zhou22}%
  \BibitemOpen
  \bibfield  {author} {\bibinfo {author} {\bibfnamefont {Yiqing}\ \bibnamefont
  {Zhou}}, \bibinfo {author} {\bibfnamefont {D.~N.}\ \bibnamefont {Sheng}}, \
  and\ \bibinfo {author} {\bibfnamefont {Eun-Ah}\ \bibnamefont {Kim}},\
  }\bibfield  {title} {\enquote {\bibinfo {title} {Quantum phases of transition
  metal dichalcogenide moir\'e systems},}\ }\href {\doibase
  10.1103/PhysRevLett.128.157602} {\bibfield  {journal} {\bibinfo  {journal}
  {Phys. Rev. Lett.}\ }\textbf {\bibinfo {volume} {128}},\ \bibinfo {pages}
  {157602} (\bibinfo {year} {2022})}\BibitemShut {NoStop}%
\bibitem [{\citenamefont {Morales-Dur\'{a}n}\ \emph {et~al.}(2021)\citenamefont
  {Morales-Dur\'{a}n}, \citenamefont {MacDonald},\ and\ \citenamefont
  {Potasz}}]{morales-duran_metal-insulator_2021}%
  \BibitemOpen
  \bibfield  {author} {\bibinfo {author} {\bibfnamefont {Nicol\'{a}s}\
  \bibnamefont {Morales-Dur\'{a}n}}, \bibinfo {author} {\bibfnamefont
  {Allan~H.}\ \bibnamefont {MacDonald}}, \ and\ \bibinfo {author}
  {\bibfnamefont {Pawel}\ \bibnamefont {Potasz}},\ }\bibfield  {title}
  {\enquote {\bibinfo {title} {Metal-insulator transition in transition metal
  dichalcogenide heterobilayer moiré superlattices},}\ }\href {\doibase
  10.1103/PhysRevB.103.L241110} {\bibfield  {journal} {\bibinfo  {journal}
  {Physical Review B}\ }\textbf {\bibinfo {volume} {103}},\ \bibinfo {pages}
  {L241110} (\bibinfo {year} {2021})}\BibitemShut {NoStop}%
\bibitem [{\citenamefont {{Morales-Dur{\'a}n}}\ \emph
  {et~al.}(2022)\citenamefont {{Morales-Dur{\'a}n}}, \citenamefont {{Potasz}},\
  and\ \citenamefont {{MacDonald}}}]{morales22}%
  \BibitemOpen
  \bibfield  {author} {\bibinfo {author} {\bibfnamefont {Nicol{\'a}s}\
  \bibnamefont {{Morales-Dur{\'a}n}}}, \bibinfo {author} {\bibfnamefont
  {Pawel}\ \bibnamefont {{Potasz}}}, \ and\ \bibinfo {author} {\bibfnamefont
  {Allan~H.}\ \bibnamefont {{MacDonald}}},\ }\bibfield  {title} {\enquote
  {\bibinfo {title} {{Magnetism and Quantum Melting in Moir{\'e}-Material
  Wigner Crystals}},}\ }\href {\doibase 10.48550/arXiv.2210.15168} {\bibfield
  {journal} {\bibinfo  {journal} {arXiv e-prints}\ ,\ \bibinfo {eid}
  {arXiv:2210.15168}} (\bibinfo {year} {2022})},\ \Eprint
  {http://arxiv.org/abs/2210.15168} {arXiv:2210.15168 [cond-mat.str-el]}
  \BibitemShut {NoStop}%
\bibitem [{si()}]{si}%
  \BibitemOpen
  \href@noop {} {\enquote {\bibinfo {title} {See supplementary material for
  additional details related to the effect of long-ranged interactions on the
  charge-ordering in the $\rm{WM}$ insulator, a brief description of vumps
  algorithm and our imps computations, the extrapolation of the spectral gaps,
  central charge, luttinger parameters, computation of the structure factors
  and error analysis of the critical scaling analysis; includes additional
  references~\cite{NOACK1996,kiely2022,white2005,mcculloch2008infinite,rams2018,vanhecke2019,osborne2022largescale,calabrese2004,giamarchi,capello2008,jiang2012,jiang2013,kiely_conservationlaws_2022,pollmann2009,kiely_thesis}.}}\
  }\BibitemShut {NoStop}%
\bibitem [{\citenamefont {Zauner-Stauber}\ \emph {et~al.}(2018)\citenamefont
  {Zauner-Stauber}, \citenamefont {Vanderstraeten}, \citenamefont {Fishman},
  \citenamefont {Verstraete},\ and\ \citenamefont {Haegeman}}]{vumps}%
  \BibitemOpen
  \bibfield  {author} {\bibinfo {author} {\bibfnamefont {V.}~\bibnamefont
  {Zauner-Stauber}}, \bibinfo {author} {\bibfnamefont {L.}~\bibnamefont
  {Vanderstraeten}}, \bibinfo {author} {\bibfnamefont {M.~T.}\ \bibnamefont
  {Fishman}}, \bibinfo {author} {\bibfnamefont {F.}~\bibnamefont {Verstraete}},
  \ and\ \bibinfo {author} {\bibfnamefont {J.}~\bibnamefont {Haegeman}},\
  }\bibfield  {title} {\enquote {\bibinfo {title} {Variational optimization
  algorithms for uniform matrix product states},}\ }\href {\doibase
  10.1103/PhysRevB.97.045145} {\bibfield  {journal} {\bibinfo  {journal} {Phys.
  Rev. B}\ }\textbf {\bibinfo {volume} {97}},\ \bibinfo {pages} {045145}
  (\bibinfo {year} {2018})}\BibitemShut {NoStop}%
\bibitem [{\citenamefont {Vanderstraeten}\ \emph {et~al.}(2019)\citenamefont
  {Vanderstraeten}, \citenamefont {Haegeman},\ and\ \citenamefont
  {Verstraete}}]{vumps2}%
  \BibitemOpen
  \bibfield  {author} {\bibinfo {author} {\bibfnamefont {Laurens}\ \bibnamefont
  {Vanderstraeten}}, \bibinfo {author} {\bibfnamefont {Jutho}\ \bibnamefont
  {Haegeman}}, \ and\ \bibinfo {author} {\bibfnamefont {Frank}\ \bibnamefont
  {Verstraete}},\ }\bibfield  {title} {\enquote {\bibinfo {title}
  {Tangent-space methods for uniform matrix product states},}\ }\href {\doibase
  10.21468/SciPostPhysLectNotes.7} {\bibfield  {journal} {\bibinfo  {journal}
  {SciPost Phys. Lect. Notes}\ ,\ \bibinfo {pages} {7}} (\bibinfo {year}
  {2019})}\BibitemShut {NoStop}%
\bibitem [{\citenamefont {Zauner}\ \emph {et~al.}(2015)\citenamefont {Zauner},
  \citenamefont {Draxler}, \citenamefont {Vanderstraeten}, \citenamefont
  {Degroote}, \citenamefont {Haegeman}, \citenamefont {Rams}, \citenamefont
  {Stojevic}, \citenamefont {Schuch},\ and\ \citenamefont
  {Verstraete}}]{Zauner2015}%
  \BibitemOpen
  \bibfield  {author} {\bibinfo {author} {\bibfnamefont {V}~\bibnamefont
  {Zauner}}, \bibinfo {author} {\bibfnamefont {D}~\bibnamefont {Draxler}},
  \bibinfo {author} {\bibfnamefont {L}~\bibnamefont {Vanderstraeten}}, \bibinfo
  {author} {\bibfnamefont {M}~\bibnamefont {Degroote}}, \bibinfo {author}
  {\bibfnamefont {J}~\bibnamefont {Haegeman}}, \bibinfo {author} {\bibfnamefont
  {M~M}\ \bibnamefont {Rams}}, \bibinfo {author} {\bibfnamefont
  {V}~\bibnamefont {Stojevic}}, \bibinfo {author} {\bibfnamefont
  {N}~\bibnamefont {Schuch}}, \ and\ \bibinfo {author} {\bibfnamefont
  {F}~\bibnamefont {Verstraete}},\ }\bibfield  {title} {\enquote {\bibinfo
  {title} {Transfer matrices and excitations with matrix product states},}\
  }\href {\doibase 10.1088/1367-2630/17/5/053002} {\bibfield  {journal}
  {\bibinfo  {journal} {New Journal of Physics}\ }\textbf {\bibinfo {volume}
  {17}},\ \bibinfo {pages} {053002} (\bibinfo {year} {2015})}\BibitemShut
  {NoStop}%
\bibitem [{\citenamefont {Eberharter}\ \emph {et~al.}(2023)\citenamefont
  {Eberharter}, \citenamefont {Vanderstraeten}, \citenamefont {Verstraete},\
  and\ \citenamefont {L\"auchli}}]{eberharter2023}%
  \BibitemOpen
  \bibfield  {author} {\bibinfo {author} {\bibfnamefont {Alexander~A.}\
  \bibnamefont {Eberharter}}, \bibinfo {author} {\bibfnamefont {Laurens}\
  \bibnamefont {Vanderstraeten}}, \bibinfo {author} {\bibfnamefont {Frank}\
  \bibnamefont {Verstraete}}, \ and\ \bibinfo {author} {\bibfnamefont
  {Andreas~M.}\ \bibnamefont {L\"auchli}},\ }\bibfield  {title} {\enquote
  {\bibinfo {title} {Extracting the speed of light from matrix product
  states},}\ }\href {\doibase 10.1103/PhysRevLett.131.226502} {\bibfield
  {journal} {\bibinfo  {journal} {Phys. Rev. Lett.}\ }\textbf {\bibinfo
  {volume} {131}},\ \bibinfo {pages} {226502} (\bibinfo {year}
  {2023})}\BibitemShut {NoStop}%
\bibitem [{\citenamefont {Szasz}\ \emph {et~al.}(2020)\citenamefont {Szasz},
  \citenamefont {Motruk}, \citenamefont {Zaletel},\ and\ \citenamefont
  {Moore}}]{szasz2020}%
  \BibitemOpen
  \bibfield  {author} {\bibinfo {author} {\bibfnamefont {Aaron}\ \bibnamefont
  {Szasz}}, \bibinfo {author} {\bibfnamefont {Johannes}\ \bibnamefont
  {Motruk}}, \bibinfo {author} {\bibfnamefont {Michael~P.}\ \bibnamefont
  {Zaletel}}, \ and\ \bibinfo {author} {\bibfnamefont {Joel~E.}\ \bibnamefont
  {Moore}},\ }\bibfield  {title} {\enquote {\bibinfo {title} {Chiral spin
  liquid phase of the triangular lattice hubbard model: A density matrix
  renormalization group study},}\ }\href {\doibase 10.1103/PhysRevX.10.021042}
  {\bibfield  {journal} {\bibinfo  {journal} {Phys. Rev. X}\ }\textbf {\bibinfo
  {volume} {10}},\ \bibinfo {pages} {021042} (\bibinfo {year}
  {2020})}\BibitemShut {NoStop}%
\bibitem [{\citenamefont {Mishmash}\ \emph {et~al.}(2015)\citenamefont
  {Mishmash}, \citenamefont {Gonz\'alez}, \citenamefont {Melko}, \citenamefont
  {Motrunich},\ and\ \citenamefont {Fisher}}]{mishmash2015}%
  \BibitemOpen
  \bibfield  {author} {\bibinfo {author} {\bibfnamefont {Ryan~V.}\ \bibnamefont
  {Mishmash}}, \bibinfo {author} {\bibfnamefont {Iv\'an}\ \bibnamefont
  {Gonz\'alez}}, \bibinfo {author} {\bibfnamefont {Roger~G.}\ \bibnamefont
  {Melko}}, \bibinfo {author} {\bibfnamefont {Olexei~I.}\ \bibnamefont
  {Motrunich}}, \ and\ \bibinfo {author} {\bibfnamefont {Matthew P.~A.}\
  \bibnamefont {Fisher}},\ }\bibfield  {title} {\enquote {\bibinfo {title}
  {Continuous mott transition between a metal and a quantum spin liquid},}\
  }\href {\doibase 10.1103/PhysRevB.91.235140} {\bibfield  {journal} {\bibinfo
  {journal} {Phys. Rev. B}\ }\textbf {\bibinfo {volume} {91}},\ \bibinfo
  {pages} {235140} (\bibinfo {year} {2015})}\BibitemShut {NoStop}%
\bibitem [{\citenamefont {Mortensen}\ \emph {et~al.}(2005)\citenamefont
  {Mortensen}, \citenamefont {Kaasbjerg}, \citenamefont {Frederiksen},
  \citenamefont {N\o{}rskov}, \citenamefont {Sethna},\ and\ \citenamefont
  {Jacobsen}}]{mortensen2005}%
  \BibitemOpen
  \bibfield  {author} {\bibinfo {author} {\bibfnamefont {J.~J.}\ \bibnamefont
  {Mortensen}}, \bibinfo {author} {\bibfnamefont {K.}~\bibnamefont
  {Kaasbjerg}}, \bibinfo {author} {\bibfnamefont {S.~L.}\ \bibnamefont
  {Frederiksen}}, \bibinfo {author} {\bibfnamefont {J.~K.}\ \bibnamefont
  {N\o{}rskov}}, \bibinfo {author} {\bibfnamefont {J.~P.}\ \bibnamefont
  {Sethna}}, \ and\ \bibinfo {author} {\bibfnamefont {K.~W.}\ \bibnamefont
  {Jacobsen}},\ }\bibfield  {title} {\enquote {\bibinfo {title} {Bayesian error
  estimation in density-functional theory},}\ }\href {\doibase
  10.1103/PhysRevLett.95.216401} {\bibfield  {journal} {\bibinfo  {journal}
  {Phys. Rev. Lett.}\ }\textbf {\bibinfo {volume} {95}},\ \bibinfo {pages}
  {216401} (\bibinfo {year} {2005})}\BibitemShut {NoStop}%
\bibitem [{\citenamefont {Zou}\ and\ \citenamefont {Chowdhury}(2020)}]{LZDC}%
  \BibitemOpen
  \bibfield  {author} {\bibinfo {author} {\bibfnamefont {Liujun}\ \bibnamefont
  {Zou}}\ and\ \bibinfo {author} {\bibfnamefont {Debanjan}\ \bibnamefont
  {Chowdhury}},\ }\bibfield  {title} {\enquote {\bibinfo {title} {Deconfined
  metallic quantum criticality: A $u(2)$ gauge-theoretic approach},}\ }\href
  {\doibase 10.1103/PhysRevResearch.2.023344} {\bibfield  {journal} {\bibinfo
  {journal} {Phys. Rev. Res.}\ }\textbf {\bibinfo {volume} {2}},\ \bibinfo
  {pages} {023344} (\bibinfo {year} {2020})}\BibitemShut {NoStop}%
\bibitem [{\citenamefont {Zhang}\ and\ \citenamefont {Sachdev}(2020)}]{YZSS}%
  \BibitemOpen
  \bibfield  {author} {\bibinfo {author} {\bibfnamefont {Ya-Hui}\ \bibnamefont
  {Zhang}}\ and\ \bibinfo {author} {\bibfnamefont {Subir}\ \bibnamefont
  {Sachdev}},\ }\bibfield  {title} {\enquote {\bibinfo {title} {Deconfined
  criticality and ghost fermi surfaces at the onset of antiferromagnetism in a
  metal},}\ }\href {\doibase 10.1103/PhysRevB.102.155124} {\bibfield  {journal}
  {\bibinfo  {journal} {Phys. Rev. B}\ }\textbf {\bibinfo {volume} {102}},\
  \bibinfo {pages} {155124} (\bibinfo {year} {2020})}\BibitemShut {NoStop}%
\bibitem [{\citenamefont {Senthil}(2008)}]{senthil2008}%
  \BibitemOpen
  \bibfield  {author} {\bibinfo {author} {\bibfnamefont {T.}~\bibnamefont
  {Senthil}},\ }\bibfield  {title} {\enquote {\bibinfo {title} {Theory of a
  continuous mott transition in two dimensions},}\ }\href {\doibase
  10.1103/PhysRevB.78.045109} {\bibfield  {journal} {\bibinfo  {journal} {Phys.
  Rev. B}\ }\textbf {\bibinfo {volume} {78}},\ \bibinfo {pages} {045109}
  (\bibinfo {year} {2008})}\BibitemShut {NoStop}%
\bibitem [{\citenamefont {Tan}\ \emph {et~al.}(2022)\citenamefont {Tan},
  \citenamefont {Tsang},\ and\ \citenamefont {Dobrosavljevi{\'{c}}}}]{VD22}%
  \BibitemOpen
  \bibfield  {author} {\bibinfo {author} {\bibfnamefont {Yuting}\ \bibnamefont
  {Tan}}, \bibinfo {author} {\bibfnamefont {Pak Ki~Henry}\ \bibnamefont
  {Tsang}}, \ and\ \bibinfo {author} {\bibfnamefont {Vladimir}\ \bibnamefont
  {Dobrosavljevi{\'{c}}}},\ }\bibfield  {title} {\enquote {\bibinfo {title}
  {Disorder-dominated quantum criticality in moir{\'e} bilayers},}\ }\href
  {\doibase 10.1038/s41467-022-35103-w} {\bibfield  {journal} {\bibinfo
  {journal} {Nature Communications}\ }\textbf {\bibinfo {volume} {13}},\
  \bibinfo {pages} {7469} (\bibinfo {year} {2022})}\BibitemShut {NoStop}%
\bibitem [{\citenamefont {Ahn}\ and\ \citenamefont {Das~Sarma}(2022)}]{SDSdis}%
  \BibitemOpen
  \bibfield  {author} {\bibinfo {author} {\bibfnamefont {Seongjin}\
  \bibnamefont {Ahn}}\ and\ \bibinfo {author} {\bibfnamefont {Sankar}\
  \bibnamefont {Das~Sarma}},\ }\bibfield  {title} {\enquote {\bibinfo {title}
  {Disorder-induced two-dimensional metal-insulator transition in moir\'e
  transition metal dichalcogenide multilayers},}\ }\href {\doibase
  10.1103/PhysRevB.105.115114} {\bibfield  {journal} {\bibinfo  {journal}
  {Phys. Rev. B}\ }\textbf {\bibinfo {volume} {105}},\ \bibinfo {pages}
  {115114} (\bibinfo {year} {2022})}\BibitemShut {NoStop}%
\bibitem [{\citenamefont {Kim}\ \emph {et~al.}(2023)\citenamefont {Kim},
  \citenamefont {Senthil},\ and\ \citenamefont {Chowdhury}}]{SKDC23}%
  \BibitemOpen
  \bibfield  {author} {\bibinfo {author} {\bibfnamefont {Sunghoon}\
  \bibnamefont {Kim}}, \bibinfo {author} {\bibfnamefont {T.}~\bibnamefont
  {Senthil}}, \ and\ \bibinfo {author} {\bibfnamefont {Debanjan}\ \bibnamefont
  {Chowdhury}},\ }\bibfield  {title} {\enquote {\bibinfo {title} {Continuous
  mott transition in moir\'e semiconductors: Role of long-wavelength
  inhomogeneities},}\ }\href {\doibase 10.1103/PhysRevLett.130.066301}
  {\bibfield  {journal} {\bibinfo  {journal} {Phys. Rev. Lett.}\ }\textbf
  {\bibinfo {volume} {130}},\ \bibinfo {pages} {066301} (\bibinfo {year}
  {2023})}\BibitemShut {NoStop}%
\bibitem [{\citenamefont {Anderson}\ \emph {et~al.}(2004)\citenamefont
  {Anderson}, \citenamefont {Lee}, \citenamefont {Randeria}, \citenamefont
  {Rice}, \citenamefont {Trivedi},\ and\ \citenamefont {Zhang}}]{RVB}%
  \BibitemOpen
  \bibfield  {author} {\bibinfo {author} {\bibfnamefont {P~W}\ \bibnamefont
  {Anderson}}, \bibinfo {author} {\bibfnamefont {P~A}\ \bibnamefont {Lee}},
  \bibinfo {author} {\bibfnamefont {M}~\bibnamefont {Randeria}}, \bibinfo
  {author} {\bibfnamefont {T~M}\ \bibnamefont {Rice}}, \bibinfo {author}
  {\bibfnamefont {N}~\bibnamefont {Trivedi}}, \ and\ \bibinfo {author}
  {\bibfnamefont {F~C}\ \bibnamefont {Zhang}},\ }\bibfield  {title} {\enquote
  {\bibinfo {title} {The physics behind high-temperature superconducting
  cuprates: the ‘plain vanilla’ version of rvb},}\ }\href {\doibase
  10.1088/0953-8984/16/24/R02} {\bibfield  {journal} {\bibinfo  {journal}
  {Journal of Physics: Condensed Matter}\ }\textbf {\bibinfo {volume} {16}},\
  \bibinfo {pages} {R755} (\bibinfo {year} {2004})}\BibitemShut {NoStop}%
\bibitem [{\citenamefont {Lee}\ \emph {et~al.}(2006)\citenamefont {Lee},
  \citenamefont {Nagaosa},\ and\ \citenamefont {Wen}}]{LNW}%
  \BibitemOpen
  \bibfield  {author} {\bibinfo {author} {\bibfnamefont {Patrick~A.}\
  \bibnamefont {Lee}}, \bibinfo {author} {\bibfnamefont {Naoto}\ \bibnamefont
  {Nagaosa}}, \ and\ \bibinfo {author} {\bibfnamefont {Xiao-Gang}\ \bibnamefont
  {Wen}},\ }\bibfield  {title} {\enquote {\bibinfo {title} {Doping a mott
  insulator: Physics of high-temperature superconductivity},}\ }\href {\doibase
  10.1103/RevModPhys.78.17} {\bibfield  {journal} {\bibinfo  {journal} {Rev.
  Mod. Phys.}\ }\textbf {\bibinfo {volume} {78}},\ \bibinfo {pages} {17--85}
  (\bibinfo {year} {2006})}\BibitemShut {NoStop}%
\bibitem [{\citenamefont {Agterberg}\ \emph {et~al.}(2020)\citenamefont
  {Agterberg}, \citenamefont {Davis}, \citenamefont {Edkins}, \citenamefont
  {Fradkin}, \citenamefont {Van~Harlingen}, \citenamefont {Kivelson},
  \citenamefont {Lee}, \citenamefont {Radzihovsky}, \citenamefont {Tranquada},\
  and\ \citenamefont {Wang}}]{PDW}%
  \BibitemOpen
  \bibfield  {author} {\bibinfo {author} {\bibfnamefont {Daniel~F.}\
  \bibnamefont {Agterberg}}, \bibinfo {author} {\bibfnamefont
  {J.C.~S\'{e}amus}\ \bibnamefont {Davis}}, \bibinfo {author} {\bibfnamefont
  {Stephen~D.}\ \bibnamefont {Edkins}}, \bibinfo {author} {\bibfnamefont
  {Eduardo}\ \bibnamefont {Fradkin}}, \bibinfo {author} {\bibfnamefont
  {Dale~J.}\ \bibnamefont {Van~Harlingen}}, \bibinfo {author} {\bibfnamefont
  {Steven~A.}\ \bibnamefont {Kivelson}}, \bibinfo {author} {\bibfnamefont
  {Patrick~A.}\ \bibnamefont {Lee}}, \bibinfo {author} {\bibfnamefont {Leo}\
  \bibnamefont {Radzihovsky}}, \bibinfo {author} {\bibfnamefont {John~M.}\
  \bibnamefont {Tranquada}}, \ and\ \bibinfo {author} {\bibfnamefont {Yuxuan}\
  \bibnamefont {Wang}},\ }\bibfield  {title} {\enquote {\bibinfo {title} {The
  physics of pair-density waves: Cuprate superconductors and beyond},}\ }\href
  {https://doi.org/10.1146/annurev-conmatphys-031119-050711} {\bibfield
  {journal} {\bibinfo  {journal} {Annual Review of Condensed Matter Physics}\
  }\textbf {\bibinfo {volume} {11}},\ \bibinfo {pages} {231--270} (\bibinfo
  {year} {2020})}\BibitemShut {NoStop}%
\bibitem [{\citenamefont {Noack}\ \emph {et~al.}(1996)\citenamefont {Noack},
  \citenamefont {White},\ and\ \citenamefont {Scalapino}}]{NOACK1996}%
  \BibitemOpen
  \bibfield  {author} {\bibinfo {author} {\bibfnamefont {R.M.}\ \bibnamefont
  {Noack}}, \bibinfo {author} {\bibfnamefont {S.R.}\ \bibnamefont {White}}, \
  and\ \bibinfo {author} {\bibfnamefont {D.J.}\ \bibnamefont {Scalapino}},\
  }\bibfield  {title} {\enquote {\bibinfo {title} {The ground state of the
  two-leg hubbard ladder a density-matrix renormalization group study},}\
  }\href {\doibase https://doi.org/10.1016/S0921-4534(96)00515-1} {\bibfield
  {journal} {\bibinfo  {journal} {Physica C: Superconductivity}\ }\textbf
  {\bibinfo {volume} {270}},\ \bibinfo {pages} {281--296} (\bibinfo {year}
  {1996})}\BibitemShut {NoStop}%
\bibitem [{\citenamefont {Kiely}\ and\ \citenamefont
  {Mueller}(2022{\natexlab{a}})}]{kiely2022}%
  \BibitemOpen
  \bibfield  {author} {\bibinfo {author} {\bibfnamefont {Thomas~G.}\
  \bibnamefont {Kiely}}\ and\ \bibinfo {author} {\bibfnamefont {Erich~J.}\
  \bibnamefont {Mueller}},\ }\bibfield  {title} {\enquote {\bibinfo {title}
  {Superfluidity in the one-dimensional bose-hubbard model},}\ }\href {\doibase
  10.1103/PhysRevB.105.134502} {\bibfield  {journal} {\bibinfo  {journal}
  {Phys. Rev. B}\ }\textbf {\bibinfo {volume} {105}},\ \bibinfo {pages}
  {134502} (\bibinfo {year} {2022}{\natexlab{a}})}\BibitemShut {NoStop}%
\bibitem [{\citenamefont {White}(2005)}]{white2005}%
  \BibitemOpen
  \bibfield  {author} {\bibinfo {author} {\bibfnamefont {Steven~R.}\
  \bibnamefont {White}},\ }\bibfield  {title} {\enquote {\bibinfo {title}
  {Density matrix renormalization group algorithms with a single center
  site},}\ }\href {\doibase 10.1103/PhysRevB.72.180403} {\bibfield  {journal}
  {\bibinfo  {journal} {Phys. Rev. B}\ }\textbf {\bibinfo {volume} {72}},\
  \bibinfo {pages} {180403} (\bibinfo {year} {2005})}\BibitemShut {NoStop}%
\bibitem [{\citenamefont {McCulloch}(2008)}]{mcculloch2008infinite}%
  \BibitemOpen
  \bibfield  {author} {\bibinfo {author} {\bibfnamefont {I.~P.}\ \bibnamefont
  {McCulloch}},\ }\href@noop {} {\enquote {\bibinfo {title} {Infinite size
  density matrix renormalization group, revisited},}\ } (\bibinfo {year}
  {2008}),\ \Eprint {http://arxiv.org/abs/0804.2509} {arXiv:0804.2509
  [cond-mat.str-el]} \BibitemShut {NoStop}%
\bibitem [{\citenamefont {Rams}\ \emph {et~al.}(2018)\citenamefont {Rams},
  \citenamefont {Czarnik},\ and\ \citenamefont {Cincio}}]{rams2018}%
  \BibitemOpen
  \bibfield  {author} {\bibinfo {author} {\bibfnamefont {Marek~M.}\
  \bibnamefont {Rams}}, \bibinfo {author} {\bibfnamefont {Piotr}\ \bibnamefont
  {Czarnik}}, \ and\ \bibinfo {author} {\bibfnamefont {Lukasz}\ \bibnamefont
  {Cincio}},\ }\bibfield  {title} {\enquote {\bibinfo {title} {Precise
  extrapolation of the correlation function asymptotics in uniform tensor
  network states with application to the bose-hubbard and xxz models},}\ }\href
  {\doibase 10.1103/PhysRevX.8.041033} {\bibfield  {journal} {\bibinfo
  {journal} {Phys. Rev. X}\ }\textbf {\bibinfo {volume} {8}},\ \bibinfo {pages}
  {041033} (\bibinfo {year} {2018})}\BibitemShut {NoStop}%
\bibitem [{\citenamefont {Vanhecke}\ \emph {et~al.}(2019)\citenamefont
  {Vanhecke}, \citenamefont {Haegeman}, \citenamefont {Van~Acoleyen},
  \citenamefont {Vanderstraeten},\ and\ \citenamefont
  {Verstraete}}]{vanhecke2019}%
  \BibitemOpen
  \bibfield  {author} {\bibinfo {author} {\bibfnamefont {Bram}\ \bibnamefont
  {Vanhecke}}, \bibinfo {author} {\bibfnamefont {Jutho}\ \bibnamefont
  {Haegeman}}, \bibinfo {author} {\bibfnamefont {Karel}\ \bibnamefont
  {Van~Acoleyen}}, \bibinfo {author} {\bibfnamefont {Laurens}\ \bibnamefont
  {Vanderstraeten}}, \ and\ \bibinfo {author} {\bibfnamefont {Frank}\
  \bibnamefont {Verstraete}},\ }\bibfield  {title} {\enquote {\bibinfo {title}
  {Scaling hypothesis for matrix product states},}\ }\href {\doibase
  10.1103/PhysRevLett.123.250604} {\bibfield  {journal} {\bibinfo  {journal}
  {Phys. Rev. Lett.}\ }\textbf {\bibinfo {volume} {123}},\ \bibinfo {pages}
  {250604} (\bibinfo {year} {2019})}\BibitemShut {NoStop}%
\bibitem [{\citenamefont {Osborne}\ \emph {et~al.}(2022)\citenamefont
  {Osborne}, \citenamefont {McCulloch}, \citenamefont {Yang}, \citenamefont
  {Hauke},\ and\ \citenamefont {Halimeh}}]{osborne2022largescale}%
  \BibitemOpen
  \bibfield  {author} {\bibinfo {author} {\bibfnamefont {Jesse}\ \bibnamefont
  {Osborne}}, \bibinfo {author} {\bibfnamefont {Ian~P.}\ \bibnamefont
  {McCulloch}}, \bibinfo {author} {\bibfnamefont {Bing}\ \bibnamefont {Yang}},
  \bibinfo {author} {\bibfnamefont {Philipp}\ \bibnamefont {Hauke}}, \ and\
  \bibinfo {author} {\bibfnamefont {Jad~C.}\ \bibnamefont {Halimeh}},\
  }\href@noop {} {\enquote {\bibinfo {title} {Large-scale $2+1$d
  $\mathrm{U}(1)$ gauge theory with dynamical matter in a cold-atom quantum
  simulator},}\ } (\bibinfo {year} {2022}),\ \Eprint
  {http://arxiv.org/abs/2211.01380} {arXiv:2211.01380 [cond-mat.quant-gas]}
  \BibitemShut {NoStop}%
\bibitem [{\citenamefont {Calabrese}\ and\ \citenamefont
  {Cardy}(2004)}]{calabrese2004}%
  \BibitemOpen
  \bibfield  {author} {\bibinfo {author} {\bibfnamefont {Pasquale}\
  \bibnamefont {Calabrese}}\ and\ \bibinfo {author} {\bibfnamefont {John}\
  \bibnamefont {Cardy}},\ }\bibfield  {title} {\enquote {\bibinfo {title}
  {Entanglement entropy and quantum field theory},}\ }\href {\doibase
  10.1088/1742-5468/2004/06/p06002} {\bibfield  {journal} {\bibinfo  {journal}
  {Journal of Statistical Mechanics: Theory and Experiment}\ }\textbf {\bibinfo
  {volume} {2004}},\ \bibinfo {pages} {P06002} (\bibinfo {year}
  {2004})}\BibitemShut {NoStop}%
\bibitem [{\citenamefont {Giamarchi}(2003)}]{giamarchi}%
  \BibitemOpen
  \bibfield  {author} {\bibinfo {author} {\bibfnamefont {Thierry}\ \bibnamefont
  {Giamarchi}},\ }\href@noop {} {\emph {\bibinfo {title} {Quantum Physics in
  One Dimension}}}\ (\bibinfo  {publisher} {Oxford University Press},\ \bibinfo
  {address} {Oxford},\ \bibinfo {year} {2003})\BibitemShut {NoStop}%
\bibitem [{\citenamefont {Capello}\ \emph {et~al.}(2008)\citenamefont
  {Capello}, \citenamefont {Becca}, \citenamefont {Fabrizio},\ and\
  \citenamefont {Sorella}}]{capello2008}%
  \BibitemOpen
  \bibfield  {author} {\bibinfo {author} {\bibfnamefont {Manuela}\ \bibnamefont
  {Capello}}, \bibinfo {author} {\bibfnamefont {Federico}\ \bibnamefont
  {Becca}}, \bibinfo {author} {\bibfnamefont {Michele}\ \bibnamefont
  {Fabrizio}}, \ and\ \bibinfo {author} {\bibfnamefont {Sandro}\ \bibnamefont
  {Sorella}},\ }\bibfield  {title} {\enquote {\bibinfo {title} {Mott transition
  in bosonic systems: Insights from the variational approach},}\ }\href
  {\doibase 10.1103/PhysRevB.77.144517} {\bibfield  {journal} {\bibinfo
  {journal} {Phys. Rev. B}\ }\textbf {\bibinfo {volume} {77}},\ \bibinfo
  {pages} {144517} (\bibinfo {year} {2008})}\BibitemShut {NoStop}%
\bibitem [{\citenamefont {Jiang}\ \emph {et~al.}(2012)\citenamefont {Jiang},
  \citenamefont {Wang},\ and\ \citenamefont {Balents}}]{jiang2012}%
  \BibitemOpen
  \bibfield  {author} {\bibinfo {author} {\bibfnamefont {Hong-Chen}\
  \bibnamefont {Jiang}}, \bibinfo {author} {\bibfnamefont {Zhenghan}\
  \bibnamefont {Wang}}, \ and\ \bibinfo {author} {\bibfnamefont {Leon}\
  \bibnamefont {Balents}},\ }\bibfield  {title} {\enquote {\bibinfo {title}
  {Identifying topological order by entanglement entropy},}\ }\href@noop {}
  {\bibfield  {journal} {\bibinfo  {journal} {Nature Physics}\ }\textbf
  {\bibinfo {volume} {8}},\ \bibinfo {pages} {902--905} (\bibinfo {year}
  {2012})}\BibitemShut {NoStop}%
\bibitem [{\citenamefont {Jiang}\ and\ \citenamefont
  {Balents}(2013)}]{jiang2013}%
  \BibitemOpen
  \bibfield  {author} {\bibinfo {author} {\bibfnamefont {Hong-Chen}\
  \bibnamefont {Jiang}}\ and\ \bibinfo {author} {\bibfnamefont {Leon}\
  \bibnamefont {Balents}},\ }\href {\doibase 10.48550/ARXIV.1309.7438}
  {\enquote {\bibinfo {title} {Collapsing schrödinger cats in the density
  matrix renormalization group},}\ } (\bibinfo {year} {2013}),\ \Eprint
  {http://arxiv.org/abs/1309.7438} {arXiv:1309.7438} \BibitemShut {NoStop}%
\bibitem [{\citenamefont {Kiely}\ and\ \citenamefont
  {Mueller}(2022{\natexlab{b}})}]{kiely_conservationlaws_2022}%
  \BibitemOpen
  \bibfield  {author} {\bibinfo {author} {\bibfnamefont {Thomas~G.}\
  \bibnamefont {Kiely}}\ and\ \bibinfo {author} {\bibfnamefont {Erich~J.}\
  \bibnamefont {Mueller}},\ }\bibfield  {title} {\enquote {\bibinfo {title}
  {Role of conservation laws in the density matrix renormalization group},}\
  }\href {\doibase 10.1103/PhysRevB.106.235126} {\bibfield  {journal} {\bibinfo
   {journal} {Phys. Rev. B}\ }\textbf {\bibinfo {volume} {106}},\ \bibinfo
  {pages} {235126} (\bibinfo {year} {2022}{\natexlab{b}})}\BibitemShut
  {NoStop}%
\bibitem [{\citenamefont {Pollmann}\ \emph {et~al.}(2009)\citenamefont
  {Pollmann}, \citenamefont {Mukerjee}, \citenamefont {Turner},\ and\
  \citenamefont {Moore}}]{pollmann2009}%
  \BibitemOpen
  \bibfield  {author} {\bibinfo {author} {\bibfnamefont {Frank}\ \bibnamefont
  {Pollmann}}, \bibinfo {author} {\bibfnamefont {Subroto}\ \bibnamefont
  {Mukerjee}}, \bibinfo {author} {\bibfnamefont {Ari~M.}\ \bibnamefont
  {Turner}}, \ and\ \bibinfo {author} {\bibfnamefont {Joel~E.}\ \bibnamefont
  {Moore}},\ }\bibfield  {title} {\enquote {\bibinfo {title} {Theory of
  finite-entanglement scaling at one-dimensional quantum critical points},}\
  }\href {\doibase 10.1103/PhysRevLett.102.255701} {\bibfield  {journal}
  {\bibinfo  {journal} {Phys. Rev. Lett.}\ }\textbf {\bibinfo {volume} {102}},\
  \bibinfo {pages} {255701} (\bibinfo {year} {2009})}\BibitemShut {NoStop}%
\bibitem [{\citenamefont {Kiely}(2024)}]{kiely_thesis}%
  \BibitemOpen
  \bibfield  {author} {\bibinfo {author} {\bibfnamefont {Thomas}\ \bibnamefont
  {Kiely}},\ }\emph {\bibinfo {title} {Phase Transitions and Transport
  Properties in Ultracold Atom Quantum Simulators}},\ \href@noop {} {Ph.D.
  thesis},\ \bibinfo  {school} {Cornell University} (\bibinfo {year}
  {2024})\BibitemShut {NoStop}%
\end{thebibliography}%

\clearpage
\renewcommand{\thefigure}{S\arabic{figure}}
\renewcommand{\figurename}{Supplemental Figure}
\setcounter{figure}{0}
\begin{widetext}
\appendix
\begin{center}
{\bf \centering Supplementary Information for \\Continuous Wigner-Mott Transitions at $\nu=1/5$\\}
{\centering Thomas G. Kiely and Debanjan Chowdhury}
\end{center}

\section{XC2 case (ii)}
Here we consider case (ii) on the XC2 geometry, in which $t^\prime=6t$. In this case, the metallic LL is in the C2S2 phase. Adopting the same procedure as before to extract the central charge (Fig.~2g), we find that $C=0$ in the Mott phase for $t\leq t_{c1}\approx 0.01U$ and $C=4$ in the LL phase for $t\geq t_{c2}\approx 0.024U$.~\footnote{Note the scales are much smaller in these units because the bandwidth $W\approx 26t$ when $t^\prime=6t$, whereas $W=6.25t$ when $t^\prime=t$.} The peak in the central charge near $t_{c2}/U$ is due to a shifting peak in the spectrum of correlation lengths,
which is a common feature when modeling phase transitions using matrix product states~\cite{szasz2020}. We expect this feature will smooth out as $\chi$ is increased. There exists a distinct intermediate gapless phase with $C=1$ for $t_{c1}\lesssim t\lesssim t_{c2}$. The evolution of the spectral gaps in the different symmetry sectors, $(Q,S_z)$, also proceeds in an identical fashion, as shown in Fig.~2h.

The results for the spin and charge SF as $C2S2$ evolves into $C0S0$ as a function of decreasing $t/U$ are quite similar to the results for the evolution from $C1S1$ to $C0S0$. The small $k$ behavior for $\N_k,~\S_k$ as a function of $t/U$ is similar for cases (i) and (ii). The singular features at the ``$2k_F$'' wavevectors require special care, as there are two distinct $k_{F1},~k_{F2}$ associated with the different Fermi points. We find that the intermediate gapless phase exhibits a singular peak at $2(k_{F1}+k_{F2})$ that appears to diverge in the limit $\chi\rightarrow\infty$; there is no singular peak at $2(k_{F1}-k_{F2})$. This suggests that the evolution from $C2S2$ to $C1S0$ is {\it not} associated with an umklapp-driven transition, which would open a gap to total charge fluctuations
\cite{musser2022}. Instead, both LLs ($C1S1$ and $C2S2$) appear to transition into an intermediate $C1S0$ phase, which is strongly reminiscent of a Luther-Emery liquid~\cite{luther1974,NOACK1996}. This is further corroborated by the evolution of the Luttinger parameters, $K_\rho,~K_\sigma$, 
which we extract from the density and spin SF, respectively. In the intermediate $C1S0$ phase, $K_\rho>0,~K_\sigma\rightarrow0$, corresponding to gapless charge fluctuations associated with the {\it total} density.

\section{Modeling long-range interactions}
Electrons in the moir{\'e} TMD bilayer experience a long-ranged Coulomb repulsion. By placing two gates parallel to the surface of the bilayer, this interaction can be screened~\cite{xu2020correlated,ghiotto2021quantum,li2021continuous}. While a variety of gate geometries could be envisioned, here we assume the common scenario of two gates equally spaced away from the moir{\'e} bilayer by a distance $d/2$. This leads to the screened potential~\cite{xu2020correlated}
\begin{equation}
    V(r)=\frac{e^2}{4\pi\epsilon\epsilon_0}\sum_{k=-\infty}^\infty\frac{(-1)^k}{\sqrt{r^2+(kd)^2}}.
\end{equation}
For both the XC2 and YC5 geometries, we choose the experimentally-relevant value $d/a=10$ where $a$ is the moir{\'e} lattice spacing.

In general, non-exponential interactions cannot be efficiently represented as a matrix product operator, meaning that iMPS simulations require some form of truncation. This leaves the additional question of where we truncate the interactions. This can be a common problem in quasi-two-dimensional geometries, as the shorter cylindrical direction will often stabilize symmetry-broken states that are unstable in the thermodynamic limit.

In Fig.~\ref{fig:classical}, we plot the classical energy (i.e. in the limit $t/U, t/V_1\to 0$) of the two low-energy charge configurations, pictured as insets, as a function of $d/a$. The upper two plots show the energy when we keep up to third-nearest-neighbor interactions, and the lower two plots show the energy when keeping up to fourth-nearest-neighbor interactions.

For the XC2 geometry, we find that the ``zig-zag" ground state (blue) always has a lower energy than the ``aligned-pairs" state (green). Moreover, due to the constrained geometry, both of these states can lower their respective energies via superexchange when $t/U>0$. For the YC5 geometry, we find that third-nearest-neighbor interactions lead to a crossover as a function of $d/a$: for $d/a\lesssim 3$, the aligned-pairs state (blue) has a lower energy, while above that value the zig-zag state (green) is the classical ground state. For the YC5 geometry, this latter state can be thought of as spatially-separated antiferromagnetic Heisenberg chains. The crossover behavior is not desirable, as the introduction of superexchange at finite $t/U$ could lead to a transition between these charge-ordered states.

Inclusion of the fourth-nearest-neighbor interaction removes this ambiguity in favor of the aligned-pairs ground state, which agrees with classical Monte Carlo simulations of the full 2D model~\cite{xu2020correlated}. Hence, for consistency, we include up to fourth-nearest-neighbor interactions on both the XC2 and YC5 geometries.

\begin{figure}[tb]
    \centering
    \includegraphics[width=4in]{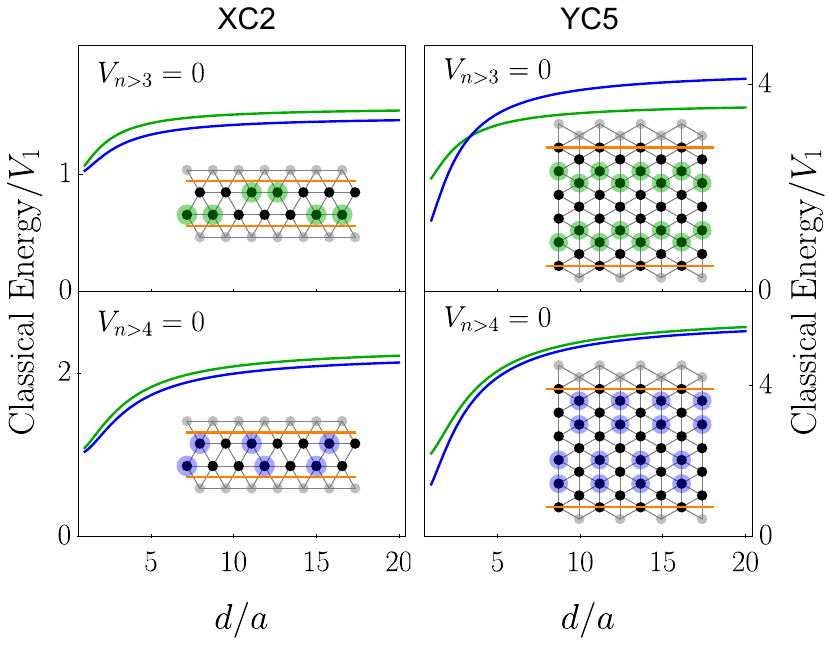}
    \caption{Classical energy ($t/U, t/V_1\to 0$) per unit cell of the two low-energy charge configurations for XC2 and YC5 geometries as a function of gate separation, $d/a$, for two possible truncations. For both XC2 and YC5, blue and green curves correspond to the blue and green charge configurations, shown as real-space cartoons in the insets.}
    \label{fig:classical}
\end{figure}
\section{Description of VUMPS}
In this section we provide a brief introduction to the variational uniform matrix product states (VUMPS) algorithm, which was developed in Ref.~\cite{vumps}. For a more extended discussion of the algorithm and notation, we refer interested readers to Refs.~\cite{vumps,vumps2,kiely2022}.

A finite MPS on an $L$-site lattice can be represented as
\begin{equation}
    |\psi\rangle =\sum_\sigma  A_1^{\sigma_1}A_2^{\sigma_2}\cdots A_L^{\sigma_L}|\sigma_1\sigma_2\cdots\sigma_L\rangle
    \label{eq:MPS}
\end{equation}
where the $A_i^{\sigma_i}\equiv [{\bf A_i}]_{s_i,s_{i+1}}^{\sigma_i}$ are rank-3 tensors that exist on physical lattice sites, enumerated by the index $i$. The variable $\sigma_i$ is taken here to represent the single-site Hilbert space of the model -- for the Fermi-Hubbard system we study here, this Hilbert space is spanned by the states $\{|\varnothing\rangle,|\uparrow\rangle,|\downarrow\rangle,|\uparrow\downarrow\rangle\}$. The state $|\sigma_1\sigma_2\cdots\sigma_L\rangle$ is understood to be a product state over the length-$L$ system. In this way, Eq.~(\ref{eq:MPS}) represents an MPS as a superposition of all possible product states. The coefficients of each product state is found by contracting the matrices $A_1^{\sigma_1}A_2^{\sigma_2}\cdots A_L^{\sigma_L}$ over the virtual indices $s_i$. The dimensions of these virtual indices are known as the ``bond dimensions" of the system. In the limit that the bond dimension is taken to infinity, Eq.~(\ref{eq:MPS}) can represent any wavefunction on a length-$L$ lattice.

The iMPS is merely the extension of Eq.~(\ref{eq:MPS}) to the limit $L\to\infty$. In order to do this, we have to make a choice that the wavefunction is periodic with a unit cell of length $n$. When contracting the iMPS, there is no longer an explicit open boundary condition. This changes what it means to take expectation values or overlaps with respect to the iMPS: rather than sequentially contracting tensors from the left or right boundary, one must consider the asymptotic behavior of the iMPS transfer matrix. For a comprehensive discussion of the properties of infinite MPS, we refer the reader to Ref.~\cite{vumps2}.

\begin{figure}[tb]
    \centering
    \includegraphics[width=3.375in]{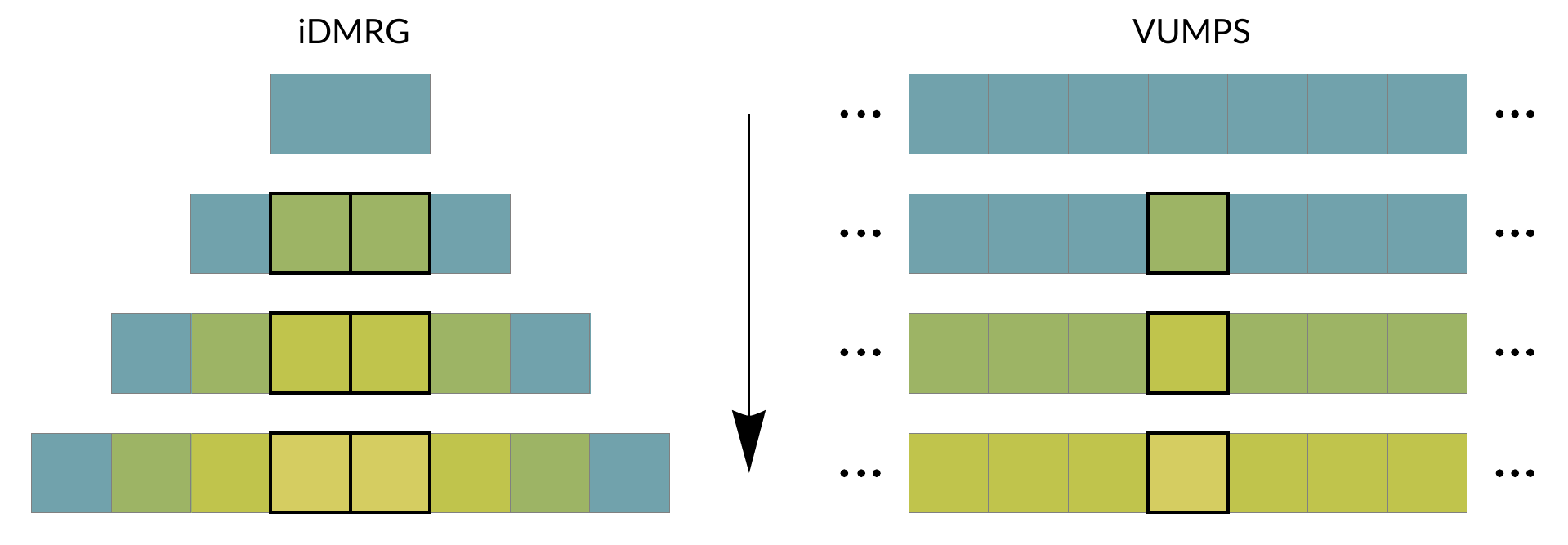}
    \caption{Cartoon representing the distinction between the iDMRG and VUMPS algorithms, reproduced from Ref.~\cite{kiely2022}. Individual squares represent the particular rank-2 or rank-3 tensors comprising an MPS. For each method, the wavefunction from the previous iteration serves as a ``bath" from which the next optimal state is chosen. While iDMRG grows a finite chain outwards by adding tensors in the center, VUMPS performs global updates after each iteration and enforces that the state be translationally-invariant.}
    \label{fig:schematic}
\end{figure}

Given this brief introduction, we can now understand the nature of the VUMPS algorithm by contrasting it with the standard way of determining ground states in the thermodynamic limit: the infinite density matrix renormalization group (iDMRG)~\cite{white2005,mcculloch2008infinite,SCHOLLWOCKreview}. The iDMRG algorithm consists of an initial finite-size MPS comprising two unit cells. One then ``grows" the initial state by adding two additional unit cells to the center of the chain and solving an eigenvalue problem to determine their lowest-energy configuration. Note that this step only modifies the additional two unit cells -- the boundary tensors remain fixed. This process is then repeated until the unit cells being added to the center have converged to a fixed point. This algorithm is illustrated pictorially on the left side of Fig.~\ref{fig:schematic}. One can then construct an iMPS out of those unit cells, which should be the ground state in the thermodynamic limit.

The VUMPS algorithm explicitly considers an iMPS at each stage of the optimization. Starting with an initial iMPS, one computes the action of the Hamiltonian on that state by computing the left and right fixed points of the MPO transfer matrix~\cite{vumps,vumps2}. Given these fixed points, one constructs an effective Hamiltonian that acts on each individual tensor of the iMPS. 
Using one of two methods~\cite{vumps}, one then updates each tensor in the unit cell and constructs a new iMPS out of the updated tensors. This is illustrated on the right half of Fig.~\ref{fig:schematic}. The process is then repeated until the variational gradient (i.e. the gradient of the variational energy with respect to the variational parameters) is reduced below a desired level. A vanishing gradient implies both that the state is truly uniform and that the wavefunction is variationally optimal. It was shown in Ref.~\cite{vumps} that the ground states obtained by iDMRG can have variational gradients as large as $\sim10^{-4}$, while VUMPS can attain considerably lower gradients.

\section{Details of the iMPS calculations}
We use the VUMPS algorithm~\cite{vumps} to find the variationally-optimal iMPS approximation to the ground state for a range of bond dimensions, $\chi$~\cite{RMP_MPS,SCHOLLWOCKreview}. In order to deduce the properties of gapless phases, we utilize finite-entanglement scaling~\cite{pollmann2009,kiely2022}. This is analogous to finite-length scaling at a critical point, where the relevant length scale is the correlation length $\xi_{\rm iMPS}$.
By studying how various measurable properties change as a function of $\xi_{\rm iMPS}$,
one can deduce the properties of  critical ground state wavefunctions.
More generally, an infinite MPS (with finite-$\chi$) is defined by a {\it spectrum} of correlation lengths. In the limit $\chi\to\infty$, one expects that this spectrum will become continuous~\footnote{Unless the state can be represented exactly as an iMPS with finite bond dimension. This case is not relevant for the present work.}. Recent works have shown that this property can be exploited to accurately extrapolate iMPS data as $\chi\to\infty$~\cite{rams2018,vanhecke2019}. For the XC2 geometry, we perform VUMPS simulations with bond dimensions $100\leq \chi\leq 1000$. For the YC5 geometry, we use bond dimensions $400\leq \chi\leq 4000$.

In our calculation, we compute the ground state at a given bond dimension using the VUMPS algorithm with single-site series-style updates~\cite{vumps}. For the XC2 geometry, we use a five-site unit cell; for the YC5 geometry, we use a ten-site unit cell. In both cases, we verify that increasing the unit cell has no effect on physical observables. At a particular bond dimension (defined as the maximum over all possible bond dimensions across the unit cell), we perform single-site VUMPS updates until the gradient has been reduced below $10^{-7}$. 

Our simulations conserve both the total particle number and total magnetization of the system. An important problem in the application of VUMPS is that the algorithm was developed with a single-site update, which does not allow one to grow the bond dimension. Furthermore, when using block-sparse (symmetry-conserving) tensors, the bond dimension must be grown carefully to ensure that relative block sizes are optimally chosen. Ref.~\cite{vumps} proposes a way of doing this via a subspace expansion, which has the benefit of preserving translational invariance in the state. For the XC2 geometry, we expand the bond dimension using this technique. For the YC5 geometry, however, we found the use of successive expansions and single-site optimizations to be cumbersome given the larger bond dimensions. For that reason, we employed a alternative integration of the time dependent variational principle (iTDVP) equations, sweeping from left to right using infinite boundary conditions. This is more natural when working with larger unit cells and can easily be extended to two-site updates, as is necessary to optimally increase the bond dimension. The technique 
is described in detail in~\cite{kiely_thesis} and 
is similar in spirit to that used in Ref.~\cite{osborne2022largescale}.

\section{Spectral gap extrapolation\label{sec:extrapolation}}
As noted in the main text, the spectrum of the iMPS transfer matrix contains all information about correlation functions in the system. In particular, any two-point correlation function can be written in the eigenbasis of the MPS transfer matrix, i.e.
\begin{equation}
\begin{split}
    \langle A_i B_j\rangle &= \left(\mathcal{J}_i(A),\left(\mathcal{T}^{i-j-1},\mathcal{J}_j(B)\right)\right)\\
    &=\sum_{n=0}^{\chi^2-1} \left(\mathcal{J}_i(A),\nu_n^R\right) \lambda_n^{i-j-1} \left(\nu_n^L,\mathcal{J}_j(B)\right)
\end{split}
\end{equation}
where $\nu^{L,R}_n$ are the left and right eigenvectors of the transfer matrix, $\lambda_n$ is the associated eigenvalue, $\mathcal{J}$ are the ``form factors" associated with the operators $A$ and $B$, and $\left(\cdots,\cdots\right)$ is an inner product with respect to the virtual degrees of freedom (i.e. the bond dimension)~\cite{Zauner2015}. Note in particular that the only term depending on the separation $|i-j|$ is $\lambda^{i-j-1}_n$. In that sense, the spectrum of the transfer matrix defines a set of correlation lengths out of which every two-point correlation function is composed.

Normalization of the wavefunction requires that the largest eigenvalue, $\lambda_0$, be equal to 1. Hence, the correlation length of the iMPS, $\xi_{\rm iMPS}$, is defined in terms of the magnitude of the second-largest transfer matrix eigenvalue:
\begin{equation}
    \xi_{\rm iMPS}\equiv -1/\ln|\lambda_1/\lambda_0| = -1/\ln|\lambda_1|.
\end{equation}
By definition, this is the longest possible correlation length of any correlation function computed with respect to the iMPS. 

\begin{figure}[tb]
    \centering
    \includegraphics[width=5.375in]{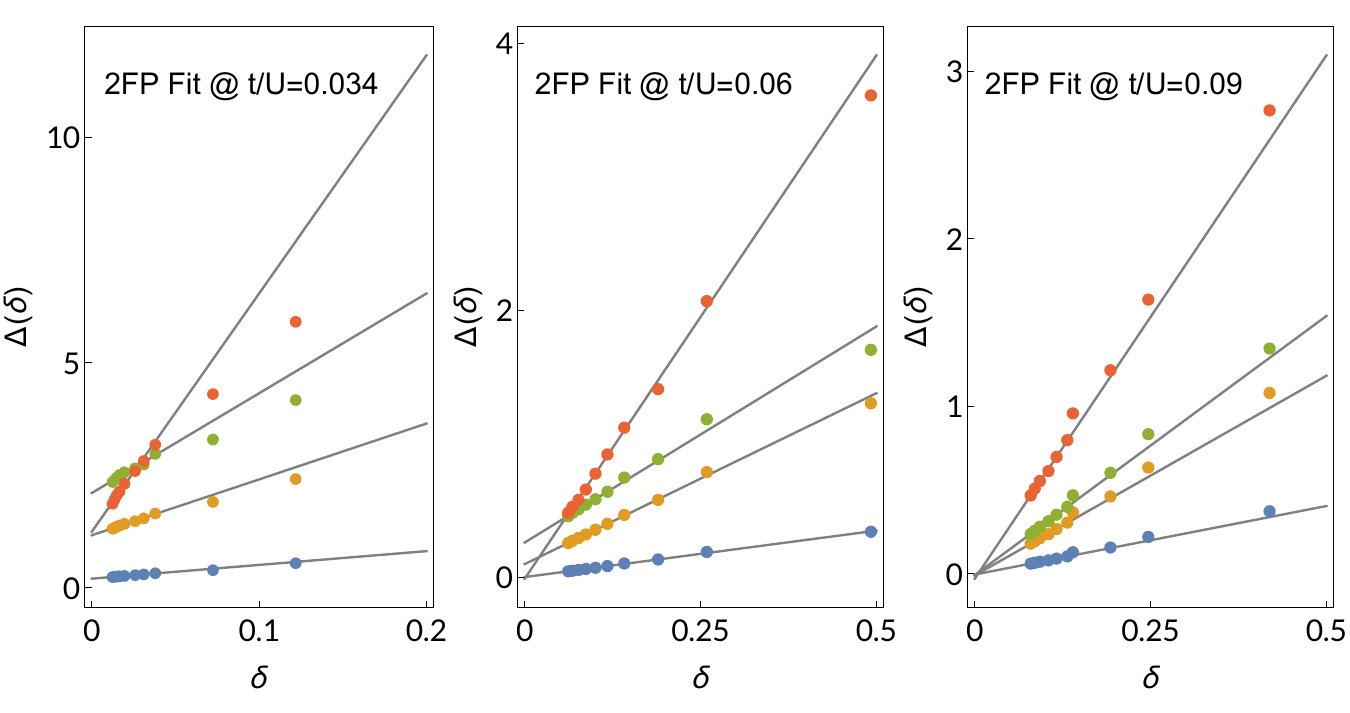}
    \caption{Example extrapolations of the two-Fermi-point XC2 data in Fig.~2 of the main text with respect to the refinement parameter, $\delta$, as described in Sec.~\ref{sec:extrapolation}. Colored dots correspond to the gaps defined in Fig.~2, while gray lines show the fit. For large bond dimensions, a linear fit with respect to the refinement parameter yields an accurate result in both the gapped and gapless phases.}
    \label{fig:extrapolate}
\end{figure}

As suggested by the analysis of Ref.~\cite{Zauner2015}, one can map the dominant eigenvalues of the transfer matrix to the low-energy spectrum of the model being studied. The justification for this is that a system with a gap $\Delta$ will have a finite correlation length $\xi\sim1/\Delta$~\cite{pollmann2009,Zauner2015}. Of course, this mapping only holds up to an overall scale factor, which was studied quantitatively in Ref.~\cite{eberharter2023}.
For this reason, we do not concern ourselves with the absolute magnitude of these correlation lengths; rather, we focus on their relative sizes and whether they scale to zero. In particular, when modeling a gapless phase, a given iMPS will always display a gap but $\lim_{\chi\to\infty}\xi_{\rm iMPS}(\chi)=0$.

We can learn even more about the state by considering the largest transfer matrix eigenvalues in different symmetry sectors. For example, if one were to compute the single-particle density matrix, $\langle c^\dagger_{i,\uparrow}c_{j,\uparrow}\rangle$, the form factors encode the fact the total particle number and total spin changed on site $i$ and changed back on site $j$. As we are conserving both particle number and spin, that means the eigenvectors $\nu_n$ that overlap with those form factors are in the sector $(Q,S_z)=(1,1/2)$, where $Q$ denotes the charge (particle number) and $S_z$ denotes the $\hat z$ magnetization. By contrast, the operator $n_{i,\uparrow}$ does not change the quantum numbers of the wavefunction, and hence the correlation function $\langle n_{i,\uparrow} n_{j,\uparrow}\rangle$ is propagated by eigenvectors in the $(Q,S_z)=(0,0)$ sector.

In order to reliably extrapolate the transfer matrix eigenvalues in the limit $\chi\to\infty$, we adopt the procedure of Refs.~\cite{rams2018,vanhecke2019}. We define a refinement parameter $\delta=\Delta_1-\Delta_0$ where $\Delta_0=1/\xi_{\rm iMPS}$ and $\Delta_1$ is the next smallest inverse correlation length. In the limit $\chi\to\infty$, $\delta$ should vanish whether one is in a gapped or a gapless phase (up to degeneracies, see generalizations in Ref.~\cite{vanhecke2019}). Hence, one can reliably extrapolate the transfer matrix eigenvalues to their asymptotic values by studying how they scale in the limit $\delta\to 0$.
In particular, we find (in agreement with Ref.~\cite{rams2018}) that the leading-order behavior scales linearly in the refinement parameter, i.e. $\Delta(\chi)=\Delta(\infty)+a\delta$ where $\{a,\Delta(\infty)\}$ and their associated errors are extracted by a linear least-squares fit. Examples of these fits in the Wigner-Mott, Luther-Emery and Luttinger liquid phases are shown in Fig.~\ref{fig:extrapolate} for the 2-Fermi-point data. 
We emphasize that this procedure yields results consistent with a direct $1/\chi\to 0$ extrapolation, but it reduces noise.

It is also worth addressing the sharp features in Fig.~2 where extrapolated transfer matrix eigenvalues appear to be negative. We believe these features to be spurious finite-bond-dimension effects. They occur at points where gaps in different symmetry sectors cross. For instance, in Fig.~\ref{fig:extrapolate} one can see that the $(Q=2,S_z=0)$ gap (in red) is large at finite-$\chi$ but extrapolates to small values. Hence, at intermediate bond dimensions in the Luther-Emery liquid phase, the $(2,0)$ eigenvalues intersect the $(0,1)$ and $(1,1/2)$ eigenvalues (which both extrapolate to finite values). These crossing points match up with the sharp features, leading us to believe that they will vanish if the calculation is pushed to larger bond dimensions.

\section{Central charge\label{sec:charge}}
When a conformally-invariant one-dimensional system is confined to a length-$L$ system, the entanglement entropy of a bipartition of the system scales as $S=\frac{C}{6}\ln(L)$ where $C$ is the conformal central charge~\cite{calabrese2004}. Although an iMPS is formally infinite, it has been established~\cite{pollmann2009,kiely2022} that its finite correlation length can be treated in a manner similar to a length cutoff. Hence, one finds that the entanglement entropy of the variational wavefunction scales as $S(\chi)=\frac{C}{6}\ln(\xi_{\rm iMPS}(\chi))$. We are able to independently compute the entanglement entropy and iMPS correlation length for each bond dimension, and hence can extract the conformal central charge via a linear regression.

For the XC2 geometry, the expected central charge of a $C\alpha S\beta$ phase (which has $\alpha$ gapless charge modes and $\beta$ gapless spin modes) is $\alpha+\beta$. For the YC5 geometry, momentum modes along the transverse direction are quantized by the periodic geometry. The expected central charge can be found from the number of times that these momentum cuts intersect the isotropic Fermi surface~\cite{szasz2020}. As shown in Fig.~1c, the $\nu=1/5$ Fermi surface is intersected by 3 momentum cuts, yielding an expected central charge of $C=6$. In Fig.~\ref{fig:cc_fit} we show some representative examples of extracting the central charge on the YC5 geometry using this linear fitting method. As shown, the fits at large bond dimension are excellent. These fits were used to extract the data in Fig.~3b.

\begin{figure}
    \centering
    \includegraphics[width=0.5\linewidth]{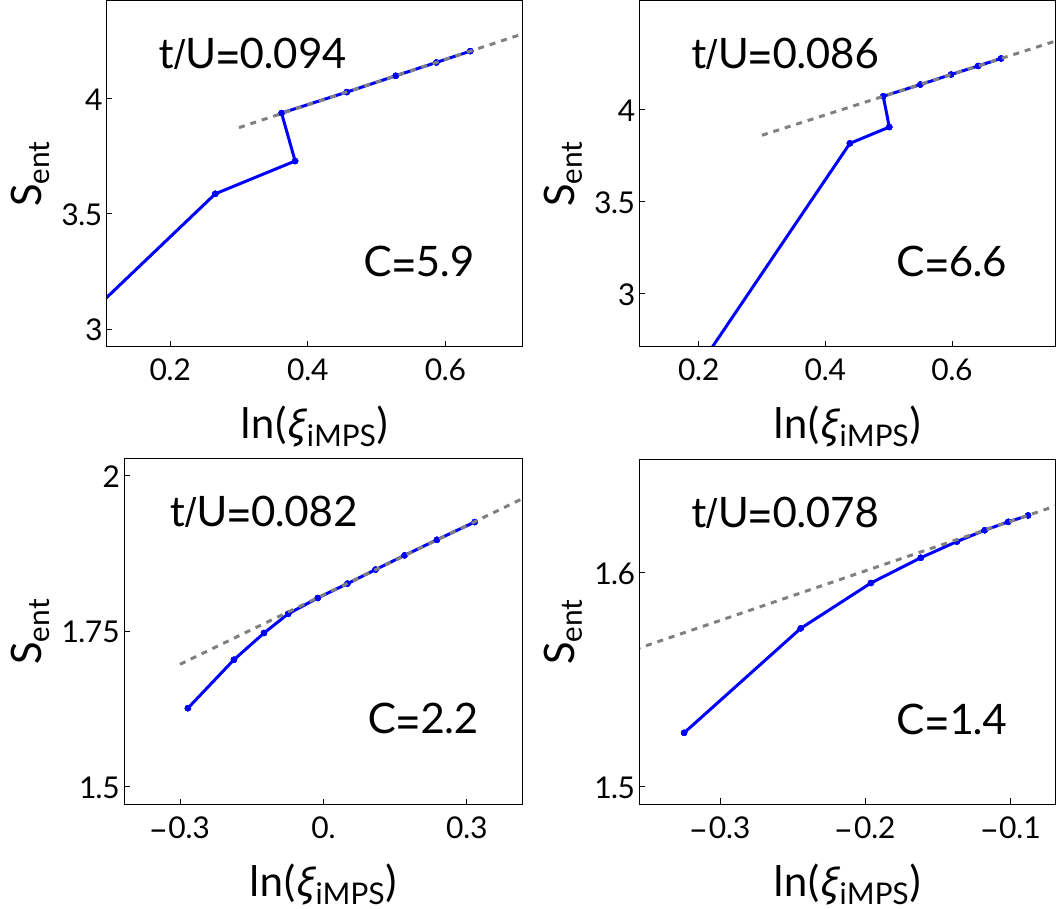}
    \caption{Plot of the linear fits to $S$ versus $\ln(\xi_{\rm iMPS})$ along with the extracted slopes for the YC5 geometry. As is evident, after the spurious low-$\chi$ features, we find very clean linear fits in the gapless regime with the expected central charge $C=6$.}
    \label{fig:cc_fit}
\end{figure}

\section{Luttinger parameters}
In one-dimensional systems, the long-wavelength behaviors of $\N_k$ and $\S_k$ are related to the charge and spin Luttinger parameters, respectively~\cite{giamarchi}. Specifically, following the definitions proposed in Ref.~\cite{mishmash2015} for spin-$1/2$ fermions on an XC2 geometry, we have
\begin{equation}
    \N_k = \frac{2K_\rho}{\pi}|k| + \mathcal{O}(k^2)\label{eq:Nlin},
\end{equation}
\begin{equation}
    \S_k = \frac{3K_\sigma}{2\pi}|k| + \mathcal{O}(k^2)\label{eq:Slin}.
\end{equation}
These definitions are formulated to capture the behavior of the two-component Luttinger liquid phase, and in particular they satisfy the non-interacting relation in which $K_\rho=K_\sigma=1$ and $\S_k=\frac{3}{4}\N_k$. Notably, when applied to the 4-Fermi-point case, these definitions yield Luttinger parameters for the {\it total} spin and charge modes, respectively~\cite{musser2022,mishmash2015}.

As an iMPS is a gapped ansatz, it will never formally satisfy the above relationships -- instead, a gapped state will have leading-order contributions ${\rm SF}(k)\propto k^2$~\cite{capello2008}. In gapless phases, however, Eqs.~(\ref{eq:Nlin}) and~(\ref{eq:Slin}) will emerge as $\chi$ is increased and $\xi_{\rm iMPS}$ vanishes. Hence, when extracting approximate Luttinger parameters numerically, it is important to keep two scales in mind: (1) Generically Eqs.~(\ref{eq:Nlin}) and~(\ref{eq:Slin}) are only satisfied below some cutoff momentum $q^{\N,\S}_{\rm cut}(t/U)$ that vanishes in the vicinity of $t_{c1}$ (for $\N_k$) or $t_{c2}$ (for $\S_k$); (2) Even when Eqs.~(\ref{eq:Nlin}) and~(\ref{eq:Slin}) hold, the correlation function of an iMPS with bond dimension $\chi$ will only faithfully represent this behavior down to momenta $q^{\N,\S}_{\rm iMPS}(\chi,t/U)\sim 1/\xi_{\rm iMPS}(\chi,t/U)$. Thus, resolving the spin and charge Luttinger parameters requires one to identify the smallest range of momenta that are faithfully represented in the iMPS correlation functions. The behavior of the Luttinger parameters in the vicinity of critical points is often rounded off, as this is a place where $q^{\N,\S}_{\rm cut}\to 0$. This behavior can be ameliorated, but never completely removed, by further increasing the bond dimension.

\begin{figure}[tb]
    \centering
    \includegraphics[width=5in]{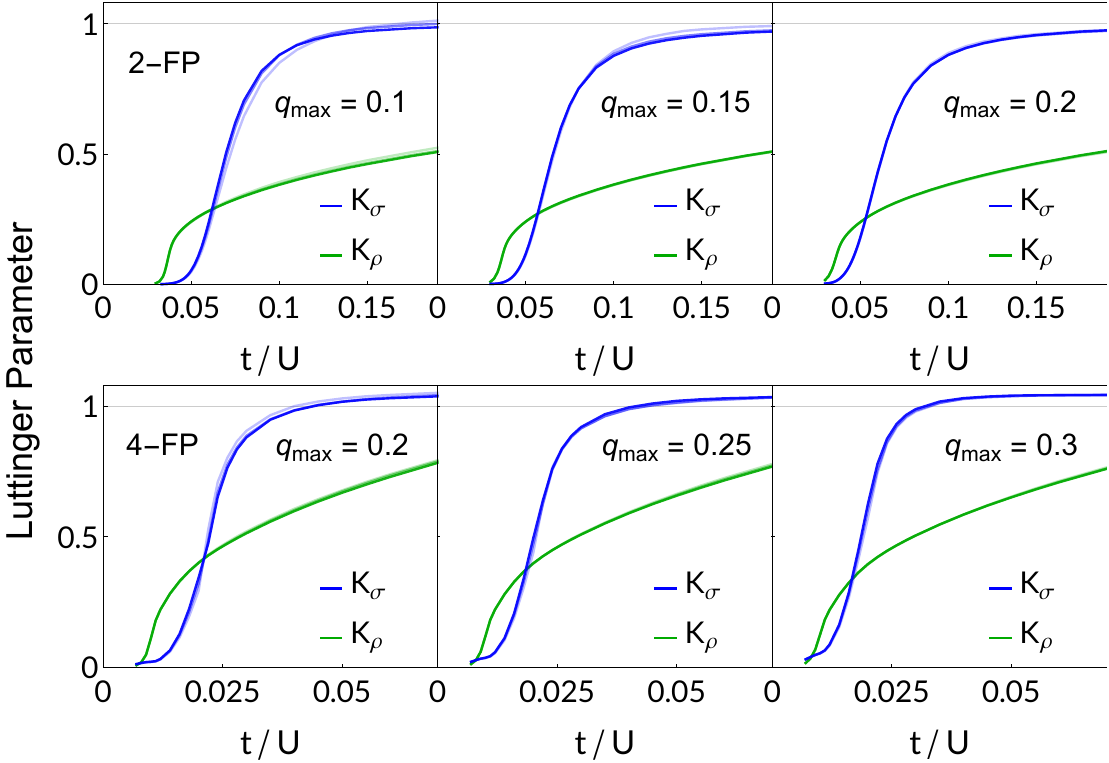}
    \caption{Plot of the Luttinger parameters $K_\rho$ and $K_\sigma$ extracted via a fit of the form $A_\mu |k|+B_\mu k^2$ to the structure factors $\N_k$ and $\S_k$ for small $k$. Top row are results for the 2-Fermi-point calculation and bottom row are for the 4-Fermi-point results. We constrain the fit to values of $|k|\leq q_{\rm max}$, with $q_{\rm max}$ labeled on each plot. Different curves correspond to different bond dimensions, with more opaque curves corresponding to a larger $\chi$.}
    \label{fig:luttinger}
\end{figure}
In Fig.~\ref{fig:luttinger} we represent this tradeoff by showing the spin and charge Luttinger parameters across the phase diagram resulting from fits of the form $A_\mu |k|+B_\mu k^2$ ($\mu=\rho,\sigma$) to $\N_k$ and $\S_k$, respectively. Each panel uses only momenta $|k|\leq q_{\rm max}$ for the fit, where $q_{\rm max}=0.1,~0.15$ and $0.2$ for the 2FP data set and $0.2,~0.25$ and $0.3$ for the 4FP data set. Each line denotes coefficients for different bond dimensions, where the more opaque curves have a larger bond dimension. 
We find that $K_\rho$ increases sharply around $t_{c1}$ and then smoothly increases as a function of $t/U$. $K_\sigma$, by contrast, remains essentially zero through $t_{c1}$ and then gradually increases, plateauing at $K_\sigma\approx 1$ around $t_{c2}$. This very gradual increase as a function of $t/U$ is expected because this latter phase transition takes place between two gapless phases~\cite{szasz2020}. The final value $K_\sigma=1$ arises in any gapless phase with ${\rm SU}(2)$ spin symmetry~\cite{giamarchi}, and hence is consistent with our expectations in the two-component Luttinger liquid phase.

\section{Structure factors\label{sec:sm_sfs}}
When computing the density structure factor, $\N_k$, we explicitly subtract off the average values of the density on respective sites. That is, the density SF is defined as
\begin{equation}
    \N_k=\frac{1}{5N^*}\sum_{j=1}^5\sum_{l=-N^*}^{N^*}e^{ilk}\langle (n_j-\langle n_j\rangle)(n_{l+j}-\langle n_{l+j}\rangle) \rangle
\end{equation}
where $N^*$ is the maximum displacement included (for the plots in the main text, $N^*=10000$). The reason for this subtraction is to remove spurious features of the gapless states: we find that all states across the phase diagram exhibit translational symmetry breaking of some form, but this vanishes asymptotically in the gapless phases. There are no such Bragg peak contributions to the spin structure factor, so we simply define it as 
\begin{equation}
    \S_k=\frac{1}{5N^*}\sum_{j=1}^5\sum_{l=-N^*}^{N^*}e^{ilk}\langle {\vec S}_j\cdot {\vec S}_{l+j}\rangle
\end{equation}
where the spin operators are given by ${\vec S}_j=(1/2)\sum_{\alpha,\beta}c^\dagger_{j,\alpha}{\vec \sigma}_{\alpha,\beta}c_{j,\beta}$.

We can study the Bragg peak contributions to $\N_k$ independently by Fourier transforming the density profile $\bar n_i\equiv\langle n_i\rangle$. As we use a five-site unit cell when studying the XC2 geometry, reflection symmetry 
dictates that there are 3 non-trivial ``Bragg peak" contributions: $\bar n_{k=0}$, $\bar n_{k=\pm 2k_F}$, and $\bar n_{k=\pm 4k_F}$. The $k=0$ component is simply the average density per site, and hence remains fixed throughout the phase diagram. The finite-$k$ components, by contrast, are not fixed -- a finite value of these peaks corresponds to long-range charge order at the associated wavevectors.

\begin{figure}[tb]
    \centering
    \includegraphics[width=4in]{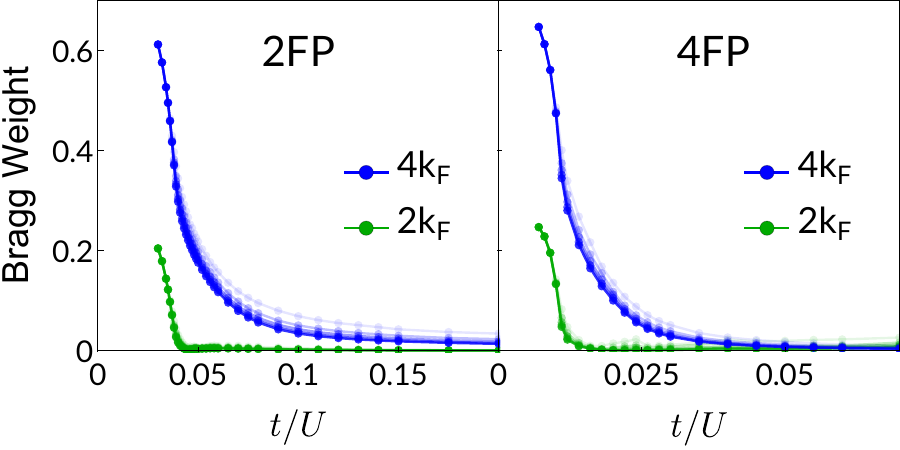}
    \caption{Bragg weights as a function of $t/U$ for the 2-Fermi-point (2FP) and 4-Fermi-point (4FP) systems. Results for finite-$\chi$ are shown, with more opaque data corresponding to larger bond dimensions. In both cases, we find that the $2k_F$ Bragg peaks vanish continuously at $t_{c1}$ while the $4k_F$ peaks persist until $t_{c2}$.}
    \label{fig:bragg}
\end{figure}

In Fig.~\ref{fig:bragg} we plot the ``Bragg weight", which we define as $|\bar n_{k}|$, at the two finite wavevectors across the phase diagram. We present the data for a variety of bond dimensions $100\leq\chi\leq 500$, where more opaque points correspond to larger bond dimensions. For both the 2-Fermi-point and 4-Fermi-point systems, the intermediate phase corresponds to a vanishing $2k_F$ Bragg peak and a persistent $4k_F$ peak. 

As noted in the main text, we use the lack of incommensurate peaks in the 4-Fermi-point $\N_k$ as an indication that the \textit{total} charge mode remains gapless in the intermediate phase, rather than the relative charge mode \cite{musser2022}. These singularities can be difficult to identify from Fig.~2 directly, so in Fig.~\ref{fig:derivativesN} we plot the derivative $d\N_k/dk$ for the same three points in the phase diagram. We also plot the derivatives $d\S_k/dk$ at the same points in Fig.~\ref{fig:derivativesS}.

\begin{figure}[tb]
    \centering
    \includegraphics[width=6in]{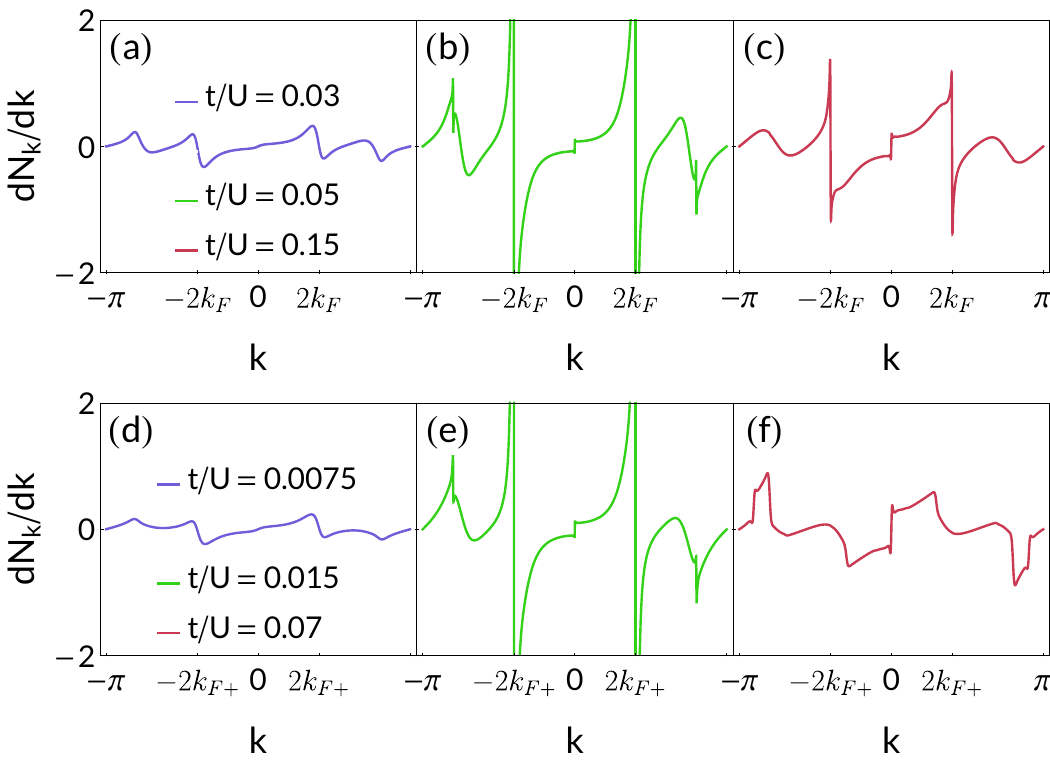}
    \caption{Derivatives $d\N_k/dk$ for the density SFs shown in Fig.~2. Panels (a)-(c) correspond to the 2-Fermi-point data while panels (d)-(f) correspond to 4-Fermi-point data. Notably, while the rightmost panel shows structure in $\N_k$ at a variety of incommensurate wavevectors, the intermediate phase shows peaks only at $2k_{F+}$ and $4k_{F+}$. This indicates that the total charge mode remains gapless.}
    \label{fig:derivativesN}
\end{figure}

\begin{figure}[tb]
    \centering
    \includegraphics[width=6in]{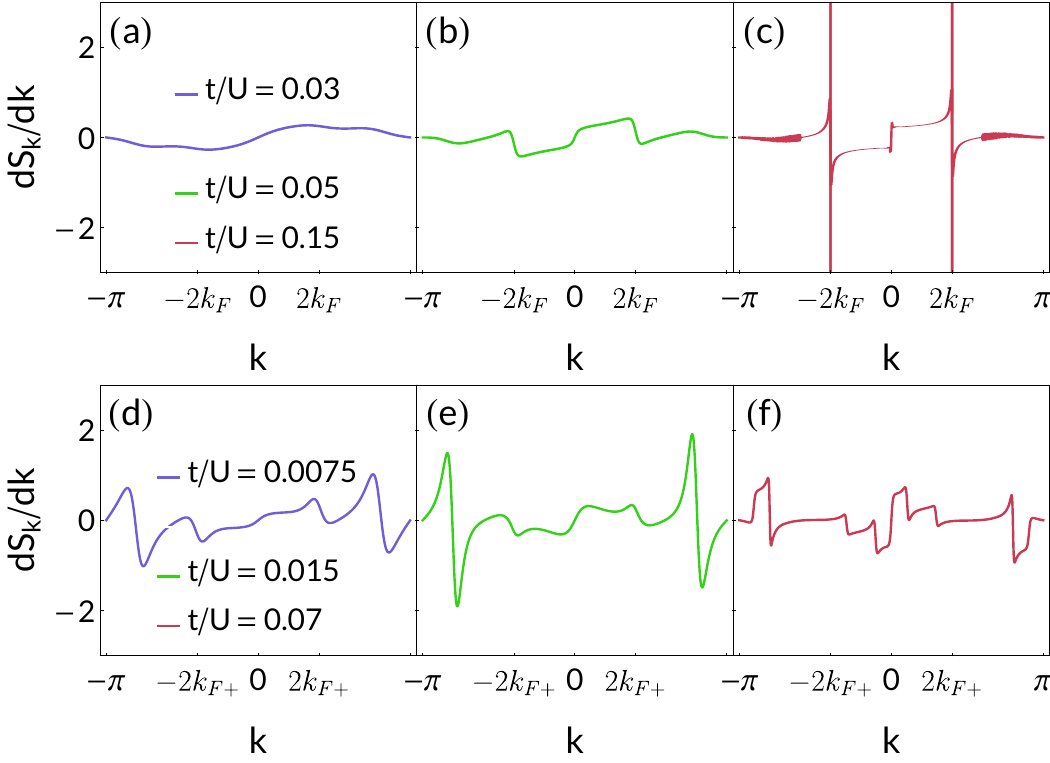}
    \caption{Derivatives $d\S_k/dk$ for the spin SFs shown in Fig.~2. Panels (a)-(c) correspond to the 2-Fermi-point data while panels (d)-(f) correspond to 4-Fermi-point data. As in Fig.~\ref{fig:derivativesN}, the rightmost panel shows structure in $\S_k$ at a variety of incommensurate wavevectors. The intermediate phases, however, do not host any sharp peaks. Note that the peak in panel (e) appears prominent but does not sharpen as $\chi$ is increased, indicating that it does not correpond to a gapless mode.}
    \label{fig:derivativesS}
\end{figure}

Of note is the fact that the 2-FP and 4-FP data are indistinguishable in the WM and intermediate phases, while in the metallic phase the 4-FP data has considerably more structure. These additional peaks occur at other (incommensurate) momenta for which 2-particle scattering events can occur between any of the 4 Fermi points.

Additionally, to support our claims that ${\rm SF}\sim|k|$ in phases where the associated sector is gapless, in Fig.~\ref{fig:zoom} we show a zoom-in to small-$k$ for the structure factors shown in Figs.~2,~\ref{fig:derivativesN} and~\ref{fig:derivativesS}, on both a linear and a log-log scale. Colors are defined to be consistent with the previous graphs. While an MPS can never have proper ${\rm SF}\sim|k|$ behavior, we see that the gapless phases exhibit approximately linear behavior up to a certain momentum scale, below which it crosses over to quadratic behavior.  In particular, note that the red $\S_k$ curves exhibit this feature for both the 2FP and 4FP data, while the green $\S_k$ curves show pronounced quadratic behavior. This is indicative of the spin gap in the Luther-Emery liquid phase. By contrast, both the green and red $\N_k$ curves are linear for the 2FP and 4FP data points.

In Fig.~\ref{fig:zoom}, note that the $\S_k$ plots in the FL regime display a finite $k=0$ value. This spurious symmetry-breaking arises due to the tendency of finite-$\chi$ MPS to break symmetries in order to reduce entanglement entropy~\cite{jiang2012,jiang2013,kiely_conservationlaws_2022}. We observe this effect in both 2 and 4-Fermi-point systems on the XC2 geometry (the symmetry-breaking is much smaller in the 2FP case), and we verify that $\S_{k=0}$ vanishes as $\chi\to\infty$.

\begin{figure}[tb]
    \centering
    \includegraphics[width=17.8cm]{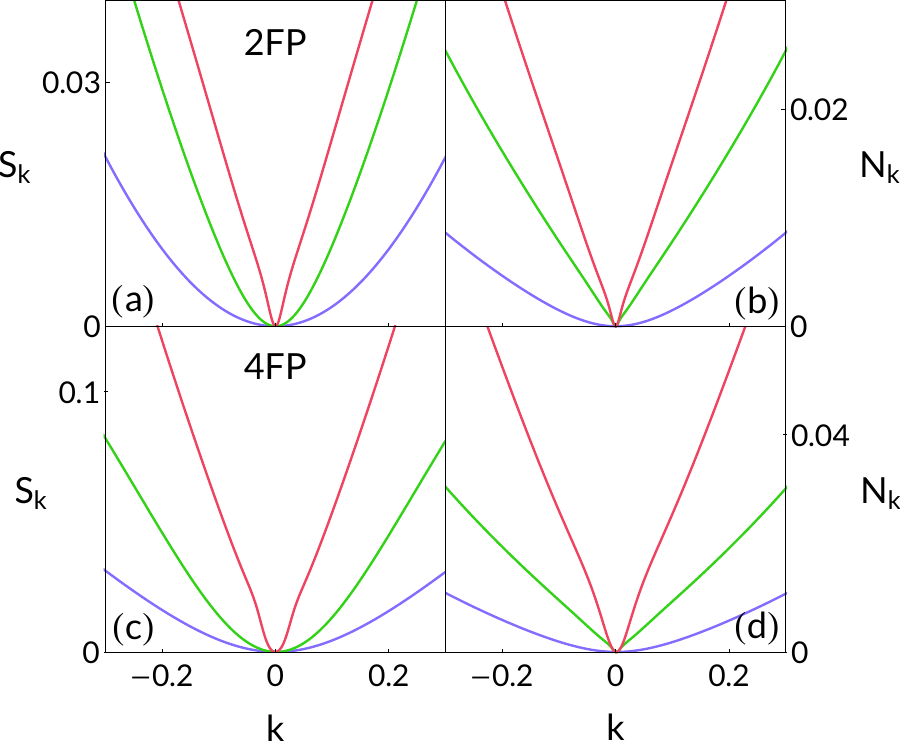}
    \caption{(a-d) Zoom in on the representative SFs (shown in Figs.~2,~\ref{fig:derivativesN} and~~\ref{fig:derivativesS}, with colors defined accordingly) near $k=0$. This shows more clearly the linear vs. quadratic distinction, elucidating the structures that were noted in the main text. Spurious symmetry-breaking of $\S_{k}$ is discussed in Sec.~\ref{sec:sm_sfs}. (e-h) Same plots on a log-log scale, with gray dashed lines as representative guides to the eye denoting $\sim|k|$ and $\sim k^2$ behaviors. Note that SF$\sim|k|$ is an emergent feature in an iMPS, and generically (for finite bond dimensions) the SF will cross over to $\sim k^2$ behavior at small momentum, as shown.}
    \label{fig:zoom}
\end{figure}

On the YC5 cylinder, we compute structure factors in an analogous way:
\begin{equation}
    \N_{\vec k}=\frac{1}{10}\sum_{j=1}^{10}\sum_{\vec r}e^{i\vec k\cdot\vec r}\langle (n_{\vec r_j}-\langle n_{\vec r_j}\rangle)(n_{\vec r_j+\vec r}-\langle n_{\vec r_j+\vec r}\rangle) \rangle
\end{equation}
\begin{equation}
    \S_{\vec k}=\frac{1}{10}\sum_{j=1}^{10}\sum_{\vec r}e^{i\vec k\cdot\vec r}\langle {\bf S}_{\vec r_j}\cdot {\bf S}_{\vec r_j+\vec r}\rangle
\end{equation}
The sums over $j$ sum over the sites in the unit cell, and $\vec r_j$ is the two-dimensional position of site $j$. We define the $\hat y$ direction as the short one (along the circumference of the cylinder). 

\section{Scaling Collapse}
The scaling collapse parameters are computed numerically by interpolating each rescaled dataset at fixed bond dimension and computing residuals of all other datasets from the interpolated curves. For the YC5 data in Fig.~3, we use only data with $\chi\geq2400$. Error bars are estimated by defining a ``thermal" probability distribution according to Ref.~\cite{mortensen2005}. We then integrate the probability distribution along the axis associated with the desired fitting parameter, producing a cumulative distribution function (CDF) for that parameter. The width of this CDF defines the error bars for our parameters. We find close agreement in the values of $t_c$, which are also displayed on Fig.~3c: $t_c(0,0)/U=0.0815(6)$,~$t_c(1,1/2)/U=0.081(1)$,~$t_c(0,1)/U=0.081(1)$,~and $t_c(2,0)/U=0.081(1)$. The remaining critical exponents are $(\zeta,\delta)_{(0,0)}=(0.17(3), -0.6(1))$, $(\zeta,\delta)_{(1,1/2)}=(0.25(5), -1.2(4))$, $(\zeta,\delta)_{(0,1)}=(0.37(8), -1.8(4))$, and $(\zeta,\delta)_{(2,0)}=(0.28(4), -1.2(3))$. 
While our error bars are too large for us to make precise statements about the exponents of this transition, we note that the ratios $\zeta/\delta$ for each of these gaps are in rough agreement with one another, with an average value of $\zeta/\delta=-0.23(8)$. This ratio controls how the gap vanishes as a function of $\chi$.

The rescaling procedure applied to the YC5 data can readily be applied to other quantum critical points. To demonstrate this, we replicate the YC5 analysis on the 2-Fermi-point XC2 data used in Fig.~2. Here we'll search for a scaling collapse in the transfer matrix eigenvalues $\Delta(Q,S_z)$, which are extrapolated in Fig.~2c.
\begin{figure}[tb]
    \centering
    \includegraphics[width=5in]{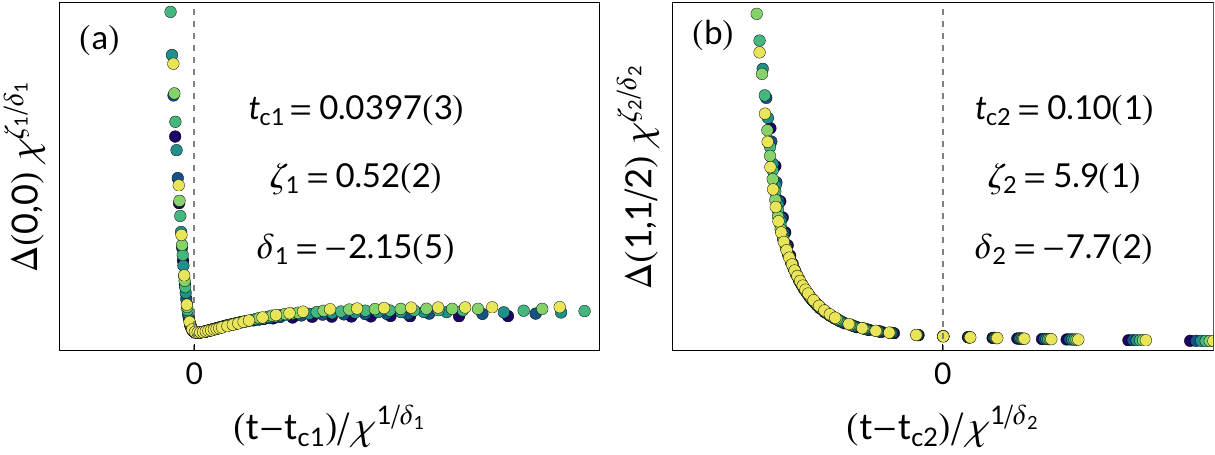}
    \caption{(a) Scaling collapse of the transfer matrix eigenvalue $\Delta(0,0)$, defined in the main text, for the 2-Fermi-point data on the XC2 lattice. We find a value of $t_{c1}$ that is in close agreement with the extrapolated value from the main text. (b) The same procedure produces a scaling collapse of the single-particle gap, $\Delta(1,1/2)$, at a value $t_{c2}\approx 0.1$. While the error bars are larger for this fit, again these results are consistent with those found from extrapolation in Fig.~2.}
    \label{fig:xc2_collapse}
\end{figure}
In Fig.~\ref{fig:xc2_collapse}a we plot the rescaled $\Delta(0,0)$, which extrapolates to zero at $t_{c1}$ as shown in Fig.~2. We assume the scaling form $\Delta(0,0;\chi)\propto \chi^{\zeta_1/\delta_1}f((t_{c1}-t)/\chi^{1/\delta_1})$ near the transition, and we extract the fit parameters $\{t_{c1}/U,\zeta_1,\delta_1\}$ by minimizing the residuals from an interpolated scaling function. We find that $(t_{c1}/U,\zeta_1,\delta_1)=(0.0397(3),0.52(2),-2.15(5))$, which are consistent with the value of $t_{c1}$ found by direct extrapolation. Fig.~\ref{fig:xc2_collapse}b shows the analogous calculation for the single-particle gap, $\Delta(1,1/2)$. There we find a scaling collapse with fitted parameters $(t_{c2}/U,\zeta_2,\delta_2)=(0.10(1),5.9(1),-7.7(2))$. Again, these are consistent with the direct-extrapolation procedure. The same method can be used to obtain error bars on $t_{c1}$ and $t_{c2}$ for the 4-Fermi-point data on an XC2 lattice. There, we find $(t_{c1}/U,\zeta_1,\delta_1)=(0.0108(3),1.55(8),-1.3(3))$ and $(t_{c2}/U,\zeta_2,\delta_2)=(0.022(2),1.0(2),-2.4(7))$.

\end{widetext}
\end{document}